\documentstyle[preprint,aps]{revtex}
\begin{document}

\title{A combined He-atom scattering and theoretical study of the 
low energy  vibrations of physisorbed monolayers of 
Xe on Cu(111) and Cu(001)} 
\author{A. \v{S}iber and B. Gumhalter\thanks{Corresponding author. 
E-mail: branko@ifs.hr}}
\address{Institute of Physics of the University, P.O. Box 304, 10001
Zagreb, Croatia}
\author{J. Braun, A.P. Graham, M. F. Bertino and J.P. Toennies}
\address{Max-Planck-Institut f\"{u}r Str\"{o}mungsforschung,
Bunsenstrasse 10, D-37073 G\"{o}ttingen, Germany}
\author{D. Fuhrmann and Ch. W\"{o}ll}
\address{Ruhr-Universit\"{a}t Bochum, Lehrstuhl f\"{u}r Physikalische 
Chemie I, Universit\"{a}tstrasse 150, 
D-44801 Bochum, Germany}

\date{\today}
\maketitle

\newcommand{\bq}{\begin{equation}} \newcommand{\eq}{\end{equation}}
\def\brho{{\hbox{\boldmath$\rho$}}}

\newpage

\begin{abstract}

The surface phonon dispersion curves of commensurate Xe monolayers on
Cu(111) and incommensurate Xe monolayers on Cu(001) surfaces have been 
measured using He atom scattering (HAS) time of flight (TOF)
spectroscopy. The TOF spectra are interpreted by combining  quantum 
scattering calculations with the dynamical matrix description of the
surface vibrations. 
Both a vertically polarized Einstein-like 
mode and another, acoustic-like mode of dominantly longitudinal
character, are identified. The latter mode is characterized by the
presence and absence of the zone center frequency gap in the
commensurate and incommensurate adlayers, respectively. The microscopic
description of the TOF spectral intensities is based on the extensive 
theoretical studies of the interplay of the phonon dynamics, 
projectile-surface potentials, multi-quantum interference and
projectile recoil, and their effect on the HAS spectra. Both single and
multi-quantum spectral features observed over a wide range of He atom 
incident energies and substrate temperatures are successfully explained
by the theory.

\end{abstract}

\newpage

\section{Introduction}
\label{sec:intro}

In the past, investigations of physisorption potentials for particles
on solid substrates as well as the potentials describing interadsorbate
interactions were mainly motivated by the experimental studies of
adsorption and phase transitions at surfaces. Several new developments
have recently stimulated renewed interest. 
First, the possibility to manipulate and displace single atoms and
molecules with low-temperature scanning tunneling
microscopes \cite{Eigler,Rieder} has not only added to our
understanding of adsorbate-substrate
interactions but also underlined the need for more detailed microscopic
data. Second, the recent progress in understanding sliding friction 
\cite{PerssonTosatti} calls for more precise
information on the
potential energy surfaces and energy dissipation to the
substrate \cite{WittePRL}. Although physisorption energies can be
determined in a rather straightforward fashion
using thermal desorption spectroscopy (TDS), information on the
interparticle potentials can be only 
extracted indirectly, either through precise analyses of the shape of 
TDS-peaks or by
detailed investigations of phase transitions within the adsorbed
adlayers \cite{Kreuzer}.

As the interplay between adsorption and interadsorbate potentials
determines the low energy dynamics of adlayers, studies of physisorbed 
rare gas monolayers are of special interest in the above discussed context 
because of their allegedly simple vibrational properties. In the present 
paper we describe a determination of the parameters of these potentials on 
the prototype commensurate and incommensurate physisorbed monolayers of Xe 
on Cu(111) and Cu(001), respectively, by using He atom scattering (HAS) time 
of flight (TOF) spectroscopy to measure the dispersion of the low-energy 
adlayer-modes in combination with a theoretical analysis of the
scattering data. 
This provides direct information on
the force constants which couple the adsorbates to the substrate, the 
effective 
corrugation of the
potential energy surface (PES) describing the motion of the particles on the 
substrate, and the force constants
coupling the adjacent adsorbates. 
Although the first HAS-TOF measurements on rare gas overlayers were
carried out 
almost twenty years ago, these early experiments were only able to detect the 
vertically polarized Einstein-like S-mode \cite{Mason:Williams,Sibener}. Only 
recently was an additional mode found for Xe on Cu(110) \cite{Zeppenfeld} and 
shortly thereafter also for Xe on Cu(001) \cite{Xe001}, Cu(111) 
\cite{BraunPRL} and NaCl(001) \cite{Gerlach} surfaces. 
This new mode was interpreted as being due to low-energy longitudinally
polarized 
motion of the adsorbates. However, simple model calculations suggest a 
significant 
discrepancy between the fits based on this assignment
\cite{Xe001,BraunPRL} and 
empirical
potential models used so far to model the adsorbate-adsorbate and 
adsorbate-substrate 
interactions. This has created some controversy in the literature 
\cite{Bruch} which 
calls for a
critical reassessment of the HAS data on the low energy vibrations 
of physisorbed Xe monolayers.

In this paper we present the analysis of the experimental data obtained
for a commensurate
${\rm(\sqrt{3}\times\sqrt{3})R30^{\circ}}$ Xe adlayer adsorbed on
Cu(111) and a similar 
system, viz. the
hexagonal incommensurate
 Xe adlayer adsorbed on Cu(001). Measurements are reported on the 
dispersion curves and 
the excitation probabilities of two adlayer induced modes, the
vertically polarized 
Einstein-like phonon branch over the
entire surface Brillouin zone (SBZ) and the new in-the-surface plane 
polarized acoustic-like 
mode over the main part of the SBZ. In
particular, for the latter mode in Xe/Cu(111) it was possible to
measure the zone-center phonon 
gap which provides particularly important
information on the Xe-substrate potential energy surface, and to
compare this result to the 
recently published data
on the incommensurate Xe-overlayer adsorbed on Cu(001)\cite{Xe001}. 
The first, tentative 
assignment of the modes observed in both systems was made by combining 
the symmetry selection 
rules applicable to the excitation of phonons of different polarization
in the single phonon 
scattering regime with the results of 
lattice-dynamical
analysis of the adlayers. 
This analysis showed that a good agreement between theoretical and 
experimental dispersion 
curves for Xe-induced modes could be obtained only with intralayer
force constants derived 
from the Xe-Xe pair interaction unexpectedly softer than the one
deduced from accurate gas 
phase potentials \cite{HFD-B2} or bulk phonon data \cite{Dove}. 
As the use of this procedure 
was recently questioned \cite{Bruch}, further progress in understanding
the HAS data for 
monolayers of Xe on Cu(111) and Cu(001) necessitates resolving this 
controversy by carrying 
out a comprehensive and consistent analysis of the measured spectral 
features. Here we 
demonstrate that additional support to the discussed assignments, and 
in particular of the 
dominantly longitudinal character of the
observed acoustic mode, can be obtained by carefully examining the 
scattering intensities 
in the HAS TOF spectra. To this end we have performed detailed
calculations of the absolute 
and relative excitation probabilities of the adlayer localized modes 
in the single and 
multiphonon scattering regime using a recently developed theoretical approach
\cite{HAS,comment} and compared them with experimental HAS TOF intensities. 
A good agreement of the results of calculations with the   experimental data 
provides a strong and convincing argument in favor of the present 
assignments and 
therefore of the strong modification of the monolayer force constants. 
It thereby 
points to some peculiar characteristics of the Xe-Xe  interaction in 
adlayers of 
monoatomic thickness which still await a microscopic interpretation 
through detailed 
calculations of the adlayer electronic and structural properties. 

The present article starts with a brief description of the experimental
procedure and the 
preparation of Xe monolayers outlined in Sec. \ref{sec:exptl}. The 
experimental results are 
presented in Sec. \ref{sec:expresul} where a tentative assignment and 
interpretation of the 
modes observed in HAS from Xe monolayers on Cu(111) and Cu(001) is 
discussed. The theoretical 
model used to interpret the HAS TOF intensities measured both in the 
single and multiphonon 
scattering regimes is described in Sec. \ref{sec:theor}. 
The basic ingredients of the model relevant to the calculations of HAS 
from adlayers are 
discussed and illustrated for the example of realistic He-Xe
interaction potentials. In Sec. 
\ref{sec:exptheor} the experimental HAS TOF intensities are compared
with the theoretical 
model predictions and physical implications of the good agreement are 
discussed. Section 
\ref{sec:corr} presents a way to determine the corrugation of the 
adlayer-substrate potential 
energy surface using the present approach. Finally, Sec.
\ref{sec:concl} summarizes all the 
relevant findings and  the most important conclusions.  
Preliminary reports on some results of the present work have been 
published previously 
\cite{Xe001,BraunPRL}.

\section{Experimental}
\label{sec:exptl}

The experiments have been carried out in a high resolution  helium atom scattering
apparatus (base pressure $8\times 10^{-11}$~mbar, fixed total scattering angle $\theta_{SD}$) described 
elsewhere~\cite{BraunApp}. 
The crystal was
mounted on a home made crystal holder and the sample could be cooled
down to 40~K with liquid 
helium. The
sample temperatures were measured using a cromel-alumel thermocouple
and could be stabilized to 
within $\pm$0.5~K. The Cu(111) crystal previously oriented to within 
${\rm 0.15^{\circ}}$ was cleaned 
by cycles of Ar ion
sputtering (600~eV, ${\rm 1\mu A/cm^2}$) followed by annealing at
1100~K.  Prior to the 
experiments
described here the cleanliness and the structural quality of the
surface were checked by 
XPS and LEED.
The 99.99\% pure Xe gas was backfilled into the scattering chamber through a
leak valve to a pressure of $4\times 10^{-8}$ mbar. During the
preparation of the Xe monolayer 
structures the crystal temperature was kept at
70~K in order to avoid the formation of bi- and multilayers, which are 
stable only at 
temperatures below
68 K.  For temperatures below 68 K the He-atom TOF spectra revealed
the appearance of an additional loss, which could be assigned to
vertically polarized  vibrations of the 
second layer of Xe atoms.
At the temperature of preparation of the Xe monolayer on Cu(111) the 
commensurate-incommensurate transitions are not expected as they have
been detected to 
occur at around 47 K \cite{Jupille}.
A description of the preparation of the Xe monolayer on the Cu(001)
surface has been given 
in Refs. \cite{Xe001,Xe}

\section{Experimental results}
\label{sec:expresul}

The right hand side of Fig. \ref{nxefg1}.a shows the structure  
of the commensurate 
${\rm (\sqrt{3}\times\sqrt{3})R30^{\circ}}$ monolayer of Xe atoms 
adsorbed on Cu(111) 
surface \cite{Chestersetal,Diehl} and indicates the two principal 
directions (azimuths)
of the substrate crystal surface. The left hand side (LHS) shows 
the first SBZ of the 
substrate (dashed lines) and the two dimensional Brillouin zone of 
the adlayer (full lines).
Figure \ref{nxefg1}.b shows an angular distribution of He atoms scattered 
from ${\rm (\sqrt{3}\times\sqrt{3})R30^{\circ}}$ Xe/Cu(111) surface 
for incident wave vector
$k_{i}=9.2$ \AA$^{-1}$ ($E_{i}=45$ meV) and the substrate
temperature $T_{s}$=60 K along the $[1{\overline1}0]$-azimuth relative 
to the substrate
surface. The intensities are normalized to the specular peak height.
In addition to the (1,0) diffraction
peak, two additional, Xe
(1/3,0) and (2/3,0) diffraction
peaks of the order one-third were observed \cite{Xe}. The sharpness of 
the peaks and 
relatively low background indicate the presence
of a well-ordered, largely defect-free Xe-overlayer.

Figure \ref{nxefg2} shows an angular distribution
along the substrate [100]-azimuth 
obtained by scattering He atoms from
a monolayer of Xe atoms adsorbed
on Cu(001) for incident wave vector $k_{i}=5.25$ \AA$^{-1}$
($E_{i}=14.36$ meV) and
$T_{s}$=52 K. This distribution indicates a well defined structure. The earlier LEED 
studies at $T_{s}\geq 77$ K have identified a hexagonally ordered 
adlayer incommensurate  
with the underlying
substrate \cite{Chesters,Glachant}, so as that the $[100]$ substrate 
direction lies along 
equivalent non-high symmetry directions for the two domains of the Xe 
monolayer, one rotated 
by $30^{0}$ from the other. The geometrical structure of these domains is shown in Fig. 6 of Ref. \cite{Chesters} and 
Fig. 2 of Ref. 
\cite{Glachant}. 
The present out-of-the-high-symmetry-plane measurements indicate sharp 
diffraction peaks 
commensurate with the orientation of the Xe overlayer along the Cu(001)
[110] azimuth except 
for the two bumps at each side of the specular peak. Their presence is 
related to inelastic 
resonance processes involving the Cu(001) surface Rayleigh wave and 
intense  nondispersive 
Xe mode with polarization perpendicular to the surface, as determined
by the TOF measurements 
described below. These characteristics of He-angular distribution spectra from  Xe/Cu(001) for $T_{s}<65$ K indicate a "floating" incommensurate Xe adlayer. 
Hence, in the case of both Cu substrates the Xe adlayers may be
considered planar and periodic 
with hexagonal symmetry, irrespective
of the (in)commensurability with the underlying substrate. The
periodicity and symmetry of the 
adlayers is then reflected in their
vibrational properties.
The adlayer vibrational modes can be classified as dominantly in-plane 
polarized (longitudinal 
(L) and shear horizontal (SH)) and shear vertical (S) \cite{Woell}.

Figure \ref{nxefg3} shows typical He atom TOF spectra for the
${\rm (\sqrt{3}\times\sqrt{3})R30^{\circ}}$Xe/Cu(111) surface along 
the $[11\bar{2}]$ 
substrate azimuth (i.e. along the $\bar{\Gamma}\bar{\mbox{K}}_{Xe}$ 
direction of the 
superstructure), for $\theta_{SD}=\theta_{i}+\theta_{f}=90.5^{0}$ and three different He atom incident
energies $E_{i}$ spanning the transition from the
single to the multi-quantum
scattering regime. The TOF spectra have been converted
from flight time to energy transfer scale. Arbitrary units are used for spectral intensities on the vertical axis. The spectrum at the lowest 
incident energy 
($E_{i}$=9.9 meV) is typical of the single
phonon scattering regime and is dominated by two well defined peaks at 
$\pm$2.62 meV on the
energy loss and gain sides of the TOF spectrum, respectively. Within
the experimental
error these energies do not change in the interval beyond  $\Delta
K=0.1$ \AA$^{-1}$ in the first SBZ of the superstructure. 
In accordance with previous works on noble gas atoms adsorbed on 
other
substrates~\cite{Sibener,Comsa}, this mode is assigned to the 
excitation of
collective vibrations of Xe-atoms with a polarization vector vertical to the
surface and is designated the S-mode. The lack of dispersion indicates 
that the frequency 
of the vertically polarized phonon is mainly determined by the
adsorbate coupling to
the substrate, with only a weak coupling to adjacent adsorbates. 
Deviation from a
dispersionless behavior occurs only at the intersection
with the substrate Rayleigh-mode \cite{Harten,Comsa}.
The  energy of this
S-mode ($\hbar\omega_{S}=2.62$ meV), is slightly larger than for 
the (110)-face of
Cu ($\hbar\omega_{S}=2.5$ meV ~\cite{Zeppenfeld}). The small,
but significant deviation of 0.12 meV is consistent with  a slightly deeper
potential well for Xe on Cu(111) and Cu(001)
than on Cu(110)~\cite{Xe:pot:theory,Xe:pot:exp}.

In addition to the intense S-peaks the measured spectrum
also reveals the presence of a weak but clearly resolved feature
(labeled "L") near the elastic or zero energy loss line. The energy of 
this mode changes 
with the angle of incidence $\theta_{i}$ and thus shows dispersion.
The spectral intensity of this mode relative to that of the S-mode in each TOF measurement was found to decrease strongly
with the magnitude of its wave vector so that for the He$\rightarrow$Xe/Cu(111) system 
the corresponding 
data points could only be
obtained for parallel wavevector up to one third of the distance
between the  $\bar{\Gamma}$ 
and $\bar{\mbox{K}}_{Xe}$ points in the first Brillouin zone of the 
superstructure (interval shown in Fig. \ref{nxefg5}).
Since in the displayed TOF spectra the energy of the "L" mode is always
significantly below 
that of the lowest surface phonon of the clean 
Cu(111) surface \cite{Cu001}, this must be a pure Xe adlayer-induced
mode which cannot couple 
to the substrate for wavevectors over a wide range of the SBZ.
As for the ${\rm (\sqrt{3}\times\sqrt{3})R30^{\circ}}$ Xe/Cu(111) system
the $[11\bar{2}]$ direction has a high symmetry mirror plane  the 
vibrational modes are 
partitioned in two
orthogonal classes \cite{deWette}. Two thirds of the modes are polarized
in the sagittal plane (including the adlayer induced S- and L-modes). 
The remaining one third 
of the modes are polarized in the surface plane and normal to the
mirror plane and designated 
shear horizontal or SH-modes. The three possible adlayer induced 
orthogonal modes with the 
wavevector in the $[11\bar{2}]$ direction (c.f. Fig. \ref{nxefg1}.a) 
are thus characterized 
by either a combination of the components with S- and L-polarization 
or pure SH-polarization.
Combining the symmetry selection rules pertinent to the probabilities 
of excitation of 
in-plane phonons at ideal surfaces (c.f. Refs.
\cite{Cellirev,Santoro} and Sec. \ref{sec:theor} below) with the fact 
that the data were 
recorded in the first SBZ of the superstructure and in the sagittal
plane which coincides 
with the high symmetry plane of the Xe/Cu(111) system, the observation 
of the SH-mode under 
these experimental conditions can be ruled out. Hence, this mode is 
tentatively assigned to 
the longitudinal mode of the adlayer which is known to couple to the 
scattered He atoms under 
similar  conditions \cite{Cellirev,Santoro}.
However, as demonstrated for NaCl, the SH-modes {\em can} be excited 
along a high symmetry 
direction in
the {\em second} SBZ \cite{Silvestri}.

The other two spectra in Fig. \ref{nxefg3} demonstrate
the transition from a single to a multiphonon scattering regime as $E_{i}$
is increased. This transition takes place already at rather low
He atom incident energies due to the very low
excitation energies of the adlayer-induced S-modes whose vertical 
polarization gives rise to a
strong projectile-phonon coupling (c.f. Sec. \ref{sec:theor}). 
Although some single phonon 
features are still discernible
at incident energy $E_{i}=21.4$ meV,
both spectra are dominated by a number of uniformly spaced peaks at energies 
$\pm n\times 2.62$ meV.
For $E_{i}=45.1$ meV the true multiphonon scattering regime is reached 
because the intensity 
of the elastic peak is smaller than that of the multi-quantum S-peak
for $n=2$.

Figure \ref{nxefg4} shows three representative
He-atom time-of-flight spectra for the incommensurate Xe monolayer on
Cu(001) for three different  He atom incident energies along the 
$\langle 100\rangle$ substrate 
azimuth which lies halfway between the two high
symmetry directions of the adlayer SBZ.
In all essential aspects these spectra are similar to those shown 
in Fig. 2. They also exhibit 
strong dispersion of the "L" mode and the
multiple excitation of S-modes at energies $\pm n\times 2.71$ meV. 
In addition, the
Rayleigh mode (labeled RW) of the underlying Cu(001) substrate is also 
observed at low and
intermediate energies $E_{i}$. The RW dispersion curves of clean
Cu(111) and Cu(001)
surfaces are well known from the previous work \cite{Harten,Cu001}.

It is noteworthy that for both adlayers the multiphonon lines are all,
to within experimental error, located at integral multiples of a 
fundamental frequency $\omega_{S}$, 
2.62 meV$/\hbar$ for Xe/Cu(111) and 2.71~meV$/\hbar$ for Xe/Cu(001).
At the first sight this seems to imply a very
harmonic Xe-Cu potential since anharmonic shifts,
which are
expected to be negative,
would produce overtone energies
smaller than the corresponding multiples of the
fundamental frequency $\omega_{S}$ (c.f. Sec. \ref{sec:corr}). 
However, the theoretical 
analyses of the Xe-metal interactions
\cite{Xe:pot:theory,Xe:pot:exp,Lang} show that the 
potential is notably anharmonic but, as we show in the next sections, the
multiple spectral peaks can be explained by multiple
excitation of delocalized phonon modes which
involve the lowest harmonic states of many adatoms
rather than a single higher anharmonic state localized on a
single adatom. In this case there is no anharmonic
shift as each multiphonon excitation is distributed
over several Xe atoms in the overlayer.

From up to about a couple of hundred of TOF spectra with different 
$\theta_{i}$  the 
experimental dispersion curves were determined and are shown in Figs. 
\ref{nxefg5} and 
\ref{nxefg6}. For both 
Xe/Cu(111) and Xe/Cu(001) the
vertically polarized S-mode exhibits negligible dispersion over the 
major part of
the SBZ except at the point of avoided crossing with the substrate
RW \cite{Hall}. The most striking difference between the vibrational 
dynamics of the two 
adsorbate phases
manifests itself in the dispersion of the "L" mode.
The "L" mode for the commensurate Xe/Cu(111) structure exhibits a zone 
center gap of about
$0.4 \pm 0.1$ meV whereas for the incommensurate phase
the frequency at the zone center goes to zero linearly with the wave vector.

To further aid the assignments of the modes in the
He$\rightarrow$Xe/Cu(111) TOF spectra and 
analyze their dispersion we have carried out a full lattice dynamics
calculation of the vibrationally coupled 
${\rm (\sqrt{3}\times\sqrt{3})R30^{\circ}}$
Xe/Cu(111) system with Xe atoms placed in on-top sites on both sides 
of a 40 layer thick slab 
of substrate atoms and interadsorbate distance $d^{Xe-Xe}=4.42$ \AA.
The interaction between nearest-neighbor Cu atoms was accounted for 
by a single radial
force constant  $\beta^{Cu}=28.0$ N/m as obtained
from a fit of the bulk Cu phonon dispersion curves \cite{Ellis}.
The other parameters describing the coupling of the Xe atoms to the 
nearest neighbor Cu 
substrate atoms was fitted to the dispersion curves, which yielded a 
radial
force constant $\beta^{Xe}=3.7$ N/m and a
tangential force constant $\alpha^{Xe}=0.086$ N/m.
Assigning the longitudinal character to the observed "L" mode to 
comply with the above 
discussed symmetry selection rules, the interaction between the atoms in the adlayer could be 
described by a radial force constant $\beta^{Xe-Xe}=0.5$ N/m and 
a tangential force constant 
$\alpha^{Xe-Xe}=0$.
The results of the full calculation are shown in Fig.~\ref{nxefg9} 
in the next section and in 
Fig. \ref{nxefg5} we present only the dispersion for the surface 
projected S- and longitudinal 
modes which reproduce the experimental data very satisfactorily.  
The radial Xe-Xe force constant $\beta^{Xe-Xe}=0.5$ N/m resulting 
from this procedure is, 
however,
significantly smaller than the value predicted from the highly 
precise HFD-B2 gas-phase
potential \cite{HFD-B2}, $\beta_{HFD}^{Xe-Xe}=1.67$ N/m, which 
produces a significantly 
steeper dispersion curve for
longitudinal phonons
denoted by dash-dotted curve in Fig. \ref{nxefg5}.

In the case of incommensurate monolayer of Xe
on Cu(001) it was
only possible to set up a three-dimensional dynamical matrix 
describing the three vibrational
modes localized in the adlayer, namely the S-, L- and SH-modes 
(c.f. Refs. \cite{Sibener,Woell}).
Since the experimental data were taken halfway between the two 
high symmetry directions of the
two dimensional hexagonal Xe adlayer Brillouin zone,
the broken symmetry no longer forbids excitation of the SH-mode. 
However, since the calculated 
polarization vector of the SH-mode is nearly
perpendicular and of the L-mode nearly parallel to the present 
azimuthal direction 
\cite{Gumhalter}, the corresponding excitation probabilities of 
the SH-phonon are expected to 
be much smaller than those of the
L-phonon (c.f. Refs. \cite{Cellirev,Santoro} and Sec. \ref{sec:theor}).
Hence, as in the case of the commensurate system, a longitudinal 
polarization is assigned to 
the observed low energy adlayer-induced acoustic "L" mode also in 
the incommensurate system 
Xe/Cu(001).
The interadsorbate distance was fixed at $d^{Xe-Xe}=4.40$ \AA and the
best-fit force constants in the [001] direction are: 
$\beta^{Xe}=3.8$ N/m, 
$\beta^{Xe-Xe}=0.42$ N/m,
and $\alpha_{T}^{Xe-Xe}=0.012$ N/m,
which are similar to force constants for the Cu(111) substrate. 
These results are also presented in Fig. \ref{nxefg6} and
reproduce the experimental data very well. For comparison the
dispersion of the
L-mode calculated by
using  $\beta_{HFD}^{Xe-Xe}$ is also shown as a dash-dotted line 
and does not fit to the data. 
However, in the case of both Xe adlayers, the physical origin of 
the unexpected large
softening of the radial Xe-Xe force constants introduced to 
reconcile the symmetry requirements 
with the experimental data remains unclear.
A clue to this effect in the case of Cu(001) and 
Cu(111) surfaces may be provided by their peculiar electronic 
structure which gives rise to surface electronic states with 
corresponding electronic wave 
functions extending far across the adsorbed Xe atoms \cite{surfstate}. Alternatively, a delocalization of the electronic charge within the monolayer itself could give rise to softening of intralayer forces which then might explain the same phenomenon observed for an insulating substrate like NaCl(001) \cite{Gerlach}.

Recently, it was suggested that the observed "L"-mode
could be interpreted as a SH-mode as this would be consistent 
with a thermodynamic analysis 
of the Xe/Cu(001) system \cite{Bruch}. This has created the 
controversy in the literature 
referred to in the Introduction and additionally motivated 
the present study.
However, the model calculations of the HAS-TOF intensities 
reported in the next sections do 
not support such an interpretation.

\section{Theoretical description of H\lowercase{e} atom scattering
from X\lowercase{e} adlayers on C\lowercase{u}(001) and C\lowercase{u}(111)}
\label{sec:theor}

Since the assignments of the peaks in the experimental TOF spectra 
discussed in
the preceding section have been questioned \cite{Bruch} additional 
corroboration by
theoretical arguments is called for. To this end we have carried out 
extensive
calculations based on the recently developed fully quantum model of 
inelastic He atom-surface
scattering \cite{HAS,comment,BGL,GBL} which is especially suitable for 
scattering
from adlayers \cite{BraunPRL,Xe,VAS}.
In this model, the single and multiphonon excitation processes can be 
treated on
an equivalent footing without invoking additional quasi-classical
approximations for the scattered particle dynamics \cite{Celli,Manson}.
A detailed description
of the model was presented in Ref. \cite{HAS}, so
only its salient properties relevant to the calculations of HAS from 
adlayers are outlined here.

In view of the HAS angular distributions characteristic of the present 
Xe/Cu(111) and
Xe/Cu(001) surfaces (c.f. Figs. \ref{nxefg1} and \ref{nxefg2}), the 
static corrugation of the 
He-surface interaction potential 
will be neglected so that the Xe overlayers are assumed to be flat and 
free of defects. The 
assumption of a planar and perfectly periodic Xe adlayer justifies the 
use of the dynamical 
matrix approach in the description
of the vibrational properties of the surface in terms of phonon modes 
characterized by their 
parallel wave vector
and branch index.  The model Hamiltonian describing inelastic 
atom-surface collisions can then 
be cast in the form:

\bq
H=H_{0}^{part} + H_{0}^{ph}  + V({\bf r}),
\label{eq:H0}
\eq
where

\bq
H_{0}^{part}=\frac{{\bf p}^2}{2M}+U(z)
\label{eq:Hpart}
\eq
is the Hamiltonian describing unperturbed motion of the projectile in the
flat static potential $U(z)$ of the target. The projectile particle is
characterized by its coordinate and momentum operators ${\bf r}=(\brho,z)$
and ${\bf p}=({\bf P},p_{z})$, respectively, and mass $M$. Here
$\brho$, $z$ and ${\bf P}$, 
$p_{z}$ denote the parallel (lateral) and vertical (normal) to the
surface  components of 
${\bf r}$ and ${\bf p}$, respectively. $H_{0}^{ph}$ is the Hamiltonian 
of the unperturbed 
phonon field which can be constructed once the dispersion
and polarization
of the vibrational modes of the Xe/Cu system are known,
and
$V({\bf r})$ is the dynamic projectile-surface interaction. The distorted
waves $\langle{\bf r}\mid {\bf k}\rangle=\langle\brho,z\mid
{\bf K},k_{z}\rangle$, which are the eigenstates of $H_{0}^{part}$,
are described by the quantum numbers ${\bf K}={\bf P}/\hbar$ and
$k_{z}=p_{z}/\hbar$ denoting, respectively, the
asymptotic parallel and normal projectile wave vectors far outside the 
range of
the potential $U(z)$. The corresponding unperturbed energy of the 
projectile is then given by
$E_{\bf k}=\hbar^{2}({\bf K}^{2}+k_{z}^{2})/2M$.
Using the box normalization these eigenstates can be orthonormalized to
satisfy
$\langle{\bf k}\mid{\bf k'}\rangle=\delta_{\bf k,k'}$,  and
are except for a phase factor equal to the unperturbed incoming and
outgoing states satisfying the scattering boundary conditions
\cite{Cellirev}. Explicitly, for a flat surface we have,
in the coordinate representation,

\bq
\langle{\bf r}\mid{\bf k}\rangle=
\frac{e^{i{\bf K}\brho}\chi_{k_{z}}(z)}{\sqrt{L_{z}L_{s}^{2}}},
\label{eq:eigenstates}
\eq
where $L_{s}$ and $L_{z}$ are the quantization lengths in the parallel 
and perpendicular to the surface directions, respectively, and
$\chi_{k_{z}}(z)$ satisfies the limit
$\chi_{k_{z}}(z\rightarrow \infty)\rightarrow 2\cos(k_{z}z+\eta)$. 
A more detailed description of the distorted wave scattering formalism 
was given in 
Refs. \cite{Cellirev,Santoro,BortoLevi,BortolaniPRB,WitteRhodium}.

The theoretical description of the scattering event described by 
Hamiltonian (\ref{eq:H0}) 
is sought in terms of a scattering spectrum
$N(\Delta E,\Delta{\bf K})$, which
gives the probability density that an amount of energy  $\Delta E$ and
parallel momentum $\hbar \Delta{\bf K}$ are transferred from the He 
atom to the substrate 
phonon field.
The particular choice of these two variables is
dictated by the symmetry of the problem, the conservation of the total 
energy and, for a 
periodic surface, the conservation of the parallel momentum to within a
reciprocal lattice 
vector ${\bf G}$.
Therefore, $\Delta E$ and $\Delta{\bf K}$ completely
determine the final state of the scattered particle provided its initial
state ${\bf k_{i}}=({\bf K_{i}},k_{zi})$
is well specified.
With these prerequisites the energy and parallel momentum resolved scattering
spectrum is defined by \cite{HAS,BGL,GBL,BortoLevi,Brenig,Meyer,Brako}

\bq
N_{\bf k_{i}} (\Delta E,\Delta{\bf
K})=\lim_{t\rightarrow\infty}
\langle\Psi (t)\mid
\delta[\Delta E -(H_{0}^{ph}- \varepsilon_{i})] \delta(\hbar\Delta{\bf
K}-{\bf \hat{P}}^{ph}) \mid \Psi (t)\rangle,
 \label{eq:spectrum}
\eq
where ${\bf \hat{P}}^{ph}$ is the parallel
momentum operator of the unperturbed phonon field,  $\varepsilon_{i}$ is
the initial energy of the phonon field,
$\Delta{\bf K=K_{i}-K_{f}}$, and $\mid \Psi (t)\rangle$ is the wave
function of the entire interacting system.
For calculational convenience, the phonon
creation (annihilation) processes are assigned positive (negative)
$\Delta E$ and $\Delta{\bf K}$ since they refer to the phonon field
quantum numbers.
The corresponding experimental quantities have, however, opposite
signs because they refer to the measured changes of energy and parallel
momentum of the 
projectile. The spectrum (\ref{eq:spectrum}) is
inherently normalized to unity and hence satisfies the optical theorem.
It is also directly proportional to the experimental time-of-flight (TOF)
spectrum \cite{comment}.
Based on the translational symmetry of the system we shall in the 
following describe the 
phonon modes propagating in Xe monolayers by their
wave vector ${\bf Q}$ parallel to the surface, branch index $j$,
polarization vector ${\bf e}({\bf Q},j)$ and energy $\hbar\omega_{{\bf Q},j}$
\cite{Sibener,Woell}.
The problem of
localized modes characteristic of either mixed layers with
broken or reduced translational symmetry \cite{Kern} or of isolated
adsorbates will be treated elsewhere.

Under the conditions in which the HAS experiments were
carried out, the uncorrelated phonon exchange processes dominate over 
the correlated ones 
\cite{HAS}. Then the angular resolved scattering spectrum is accurately
represented by the 
expression:

\bq
N^{EBA}_{{\bf k_{i}},T_{s}}(\Delta E,\Delta{\bf K}) =
\int^{\infty}_{-\infty} \frac{d\tau d^{2}{\bf R}}{(2\pi\hbar)^{3}}
e^{\frac{i}{\hbar}[(\Delta E)\tau-\hbar(\Delta{\bf K}){\bf R}]}
\exp[2W^{EBA}({\bf R},\tau)-2W^{EBA}(0,0)], \label{eq:specEBA}
\eq
where $\tau$ and ${\bf R}=(X,Y)$ are auxiliary variables occurring 
after the temporal and 
spatial Fourier representation of the energy and parallel momentum 
delta-functions, 
respectively, are introduced on the RHS of expression
(\ref{eq:spectrum}) \cite{HAS}, 
$2W^{EBA}({\bf R},\tau)$ is the exponentiated Born approximation 
(EBA) expression for 
the so-called
scattering or driving function which contains all information
on uncorrelated phonon exchange processes in the atom-surface
scattering event.
Its zero point
value $2W^{EBA}(0,0)=2W^{EBA}_{T_{s}}$ gives the Debye-Waller exponent
(DWE) in
the EBA and the corresponding Debye-Waller factor (DWF),
$\exp[-2W^{EBA}_{T_{s}}]$, gives the probability of the elastically scattered
specular beam \cite{DWF}. Since in the EBA the
correlations between two subsequent phonon scattering events are
neglected, the expression for $N^{EBA}_{{\bf
k_{i}},T_{s}}(\Delta E,\Delta{\bf K})$ on the LHS of
(\ref{eq:specEBA}) must be combined with the conservation laws for the 
total energy and parallel momentum \cite{Evans-Mills}.

The EBA expression for the scattering spectrum (\ref{eq:specEBA}) holds
irrespective of the form of the projectile-phonon coupling. However, it
has been shown \cite{MansonArmand} that, for the projectile-phonon coupling
to all orders in the lattice displacements, the higher order phonon exchange
processes which involve only single phonon vertices and originate from linear
coupling give much larger contribution
to the scattering matrix than the non-linear many-phonon  processes of the
same multiplicity (c.f. Fig. 1 of Ref. \cite{MansonArmand} and Fig. 1 of Ref.
 \cite{HAS}).
Hence, in the present approach, only the linear projectile-phonon
coupling will be retained, in which case the scattering function takes 
the form \cite{HAS}:

\begin{eqnarray} 2W^{EBA}({\bf R},\tau) &=& \sum_{{\bf Q,G},j,k_{z}'} \left[
\mid {\cal V}^{{\bf
K_{i},Q+G},j}_{k_{z}',k_{zi}}(+)\mid^{2}[\bar{n}(\hbar\omega_{{\bf Q},j})+1]
e^{-i(\omega_{{\bf Q},j}\tau -{\bf (Q+G)R})}\right. +\nonumber\\ &+& \left.
\mid {\cal V}^{{\bf K_{i},Q+G},j}_{k_{z}',k_{zi}}(-)
\mid^{2}\bar{n}(\hbar\omega_{{\bf Q},j}) e^{i(\omega_{{\bf Q},j}\tau -{\bf
(Q+G)R})}\right].
\label{eq:WEBA}
\end{eqnarray}
Here $\bar{n}(\omega_{{\bf Q},j})$ is the Bose-Einstein
distribution of phonons of energy $\hbar\omega_{{\bf Q},j}$ at the substrate
temperature $T_{s}$ and the symbols

\bq
{\cal V}^{{\bf K_{i},Q+G},j}_{k_{z}',k_{zi}}(\pm)= 2\pi V^{{\bf K_{i}\mp
Q+G,K_{i}},j}_{k_{z}',k_{zi}} \delta(E_{{\bf K_{i}\mp Q+G},k_{z}'}-E_{{\bf
K_{i}},k_{zi}} \pm \hbar\omega_{{\bf Q},j})
\label{eq:calV}
\eq
denote the on-the-energy and momentum-shell one-phonon emission (+) and
absorption $(-)$ matrix elements or the probability amplitudes of inelastic
He atom-surface scattering \cite{HAS} expressed through the
corresponding off-the-energy-shell 
interaction
matrix elements $V^{{\bf K_{i}\mp
Q+G,K_{i}},j}_{k_{z}',k_{zi}}$ (see below).
In the present fully three dimensional calculations the
wave vectors of real phonons exchanged in the collision will be restricted to
the first SBZ of the superstructure (i.e. {\bf G}=0) because the major part
of the experimental data was recorded in this region.
However, for integrated quantities such as the DWF, the relevant
summations sometimes need 
to be extended beyond the first SBZ (see below).
The matrix elements given by expression (\ref{eq:calV})
are normalized to unit particle
current normal to the surface, $j_{z}=v_{z}/L_{z}=\hbar k_{z}/ML_{z}$. 
This can be easily 
verified if according to the box normalization the energy conserving
$\delta$-function in (\ref{eq:calV}) is converted into the Kronecker symbol
following the identity

\bq
2\pi\delta(E_{{\bf K\mp Q},k_{z}'}-E_{{\bf
K},k_{z}} \pm \hbar\omega_{{\bf Q},j})=\frac{L_{z}}{\hbar\sqrt{v_{z}'v_{z}}}
\delta_{k_{z}',\bar{k}_{zi}}\Theta(\bar{k}_{zi}^{2}),
\label{eq:Kronecker}
\eq
where 
$\bar{k}_{zi}^{2}=
\pm 2{\bf K_{i}\cdot Q}-{\bf Q}^{2} +k_{zi}^{2}\mp 2M\omega_{{\bf
Q},j}/\hbar$ and $\Theta(\bar{k}_{zi}^{2})$ is the step function 
restricting $\bar{k}_{zi}^{2}$
only to open scattering channels in which $\hbar k_{z}'^{2}/2M>0$.
The factor of $L_{z}$ appearing in (\ref{eq:Kronecker}) is canceled by the
factor $L_{z}^{-1}$ appearing in 
$V^{{\bf K_{i}\mp Q,K_{i}},j}_{k_{z}',k_{zi}}$ which
arises from the box normalization of the projectile wave functions (c.f. Eq.
(\ref{eq:eigenstates})). This enables a straightforward summation over
$k_{z}'$ on the RHS of expression (\ref{eq:WEBA}). The quantization 
lengths in the
parallel directions also cancel out from expression (\ref{eq:WEBA}) after the
final summation over ${\bf Q}$. The Debye-Waller exponent is then
obtained from 
Eq. (\ref{eq:WEBA}) by a
straightforward substitution ${\bf R}=0$ and $\tau=0$.

In the following we shall adopt for 
$V^{{\bf K_{i}\mp Q,K_{i}},j}_{k_{z}',k_{zi}}$
the expressions obtained by taking the matrix elements of the vibrating part
$V({\bf r})$ of the total He-Xe/Cu  potential $V_{tot}({\bf r})$ which will
be modeled by a
pairwise sum \cite{group} of atomic He-Xe interaction potentials
$v({\bf r-r_{l}})$ \cite{TT,Kleine}. Hence:

\bq
V({\bf r})=\left(V_{tot}({\bf r})\right)_{vib}=
\left(\sum_{\bf l}v({\bf r-r_{l}-u_{l}})\right)_{vib},
\label{eq:Vtot}
\eq
where ${\bf r_{l}}=({\bf \brho_{l}},z_{l})$ ranges over the
equilibrium positions of Xe atoms and
${\bf u_{l}}$ denotes the displacement of the ${\bf l}$-th Xe atom 
from equilibrium. 
The static part of $V_{tot}({\bf r})$, which is equal to the average 
of the pairwise sum 
in the brackets on the RHS of Eq.
(\ref{eq:Vtot}), is then identified with the static atom-surface potential
$U(z)$ which is included in $H_{0}^{part}$ in Eq. (\ref{eq:Hpart}). 
The thus formulated scattering theory is equally appropriate  to treat 
single and multiphonon 
scattering processes in HAS and incorporates the single phonon DWBA 
scattering theory reviewed 
in Refs. \cite{Cellirev} and \cite{Santoro} as a special limit.
 
To simplify
the numerical calculations and in particular the treatment of the 
interaction matrix 
elements and the
scattering function, we shall approximate $U(z)$ by a suitably adjusted Morse
potential: 

\bq
U(z)=D(e^{-2\alpha(z-z_{0})}-2e^{-\alpha(z-z_{0})}),
\label{eq:Morse}
\eq
where $D$, $z_{0}$ and $\alpha$ denote the potential well depth, position
of the minimum and the inverse range, respectively. This approximation
is justified in the 
range of energies of the present HAS experiments \cite{Gaspar}. The
values of these potential 
parameters appropriate to the two studied collision systems are given 
in Sec. \ref{sec:exptheor} 
in which they are explicitly quoted as the input in the  calculations 
of the scattering 
intensities for comparison with the experimental results. 

Following the findings of Ref.~\cite{Siber} that the repulsive and
attractive components of the pair potentials contribute to the vibrating
part of the total dynamic atom-surface interaction, both 
components are included in the dynamic atom-adlayer interaction. Then the
matrix elements for linear atom-phonon
coupling acquire a simple form \cite{Cellirev,Santoro}:

\bq
V_{k_{z}',k_{z}}^{{\bf K\mp Q,K},j}=
{\bf u}^{*}({\bf Q},j)\cdot {\bf F}({\bf K'-K},k_{z}',k_{z})
\delta_{\bf K',K\mp Q},
\label{eq:V}
\eq
where ${\bf u}({\bf Q},j)$ is the vector of quantized displacement of the
adlayer atoms corresponding to the  phonon mode $({\bf Q},j)$ defined below
in Eq. (\ref{eq:phonexp}). The parallel momentum conserving Kronecker symbols
$\delta_{\bf K',K\mp Q}$ arise as a result of the summation over adsorption
sites of pair potential contributions from all adsorbates in the periodic
adlayer. The matrix element of the force ${\bf F}({\bf K'-K},k_{z}',k_{z})$
exerted on the projectile by an atom in the adlayer is expressed as
\cite{Cellirev,Santoro}:

\bq
{\bf F}({\bf K'-K},k_{z}',k_{z}) =
\langle \chi_{k_{z}'}\mid (i({\bf K'-K}), -\partial/\partial z)
v({\bf K'-K},z)\mid \chi_{k_{z}}\rangle,
\label{eq:F}
\eq
where $v({\bf K'-K},z)$  is a two-dimensional Fourier transform of
$v({\bf r})$ which appears in expression (\ref{eq:F}) after taking the
matrix element between the parallel components of the projectile wave 
functions
$\langle \brho\mid {\bf K'}\rangle$ and $\langle \brho\mid {\bf K}\rangle$.
The scalar product appearing in expression (\ref{eq:V}) gives the afore
mentioned symmetry 
selection rules for excitation
amplitudes of the various one-phonon scattering processes. In
particular, for a purely 
SH-polarized phonon with a polarization vector in the surface plane and
perpendicular to the 
mode wave vector ${\bf Q}$ in the first SBZ, the scalar product in
(\ref{eq:V}) will be equal to zero, leading to a vanishing excitation 
probability. 
Expressions (\ref{eq:V}) and (\ref{eq:F}) also indicate that for small 
parallel momentum 
transfer to phonons the projectile-phonon coupling is strongest for 
vibrations with polarization 
vector perpendicular to the surface. 

The parallel or $({\bf K'-K})$-dependence  of  $v({\bf K'-K},z)$ was 
modeled  by introducing 
the cut-off wave
vector  $Q_{c}$ which gives an
approximate upper bound on the parallel momentum transfer in the 
one-phonon
exchange processes (Hoinkes-Armand effect \cite{Hoinkes}). Denoting 
${\bf K'-K=Q}$ this yields:

\bq
v({\bf Q},z)=
D(e^{-2\alpha(z-z_{0})}e^{-Q^{2}/2Q_{c}^{2}}
-2e^{-\alpha(z-z_{0})}e^{-Q^{2}/Q_{c}^{2}}).
\label{eq:v}
\eq
with the property $v({\bf Q}=0,z)=U(z)$.
The appearance of different effective cut-offs in the
$Q$-dependence of the repulsive and attractive components of $v({\bf Q},z)$
in Eq. (\ref{eq:v}) (viz. $\sqrt{2}Q_{c}$ versus $Q_{c}$) is a consequence of
the different ranges of the repulsive and attractive components of the model
pair potential $v({\bf r-r_{l}})$ consistent with expression 
(\ref{eq:Morse}) \cite{Siber}.
In the case of exponential potentials of range $1/\beta$ the value of
$Q_{c}$ is approximately given by \cite{group}:

\bq
Q_{c}=\sqrt{\frac{\beta}{z_{t}}},
\label{eq:Qc}
\eq
where $z_{t}$ is the (energy dependent) average value of the He atom turning
point in the surface potential $U(z)$. Although the numerical
evaluation of the matrix
elements (\ref{eq:F})  avoids the  explicit introduction of $Q_{c}$
\cite{Gaspar} the effect of the cut off in the parallel momentum 
transfer remains 
\cite{Tommasini}. The matrix elements described by Eq. (\ref{eq:F}) 
were all calculated by 
using the pair potentials $v({\bf r-r_{l}})$ to obtain $v({\bf Q},z)$, 
Eq. (\ref{eq:v}), with 
the  parameters corresponding to the two studied systems explicitly 
defined in the next 
section.
Figures  \ref{nxefg7}.a and \ref{nxefg7}.b illustrate the behavior 
of the one dimensional 
off-shell matrix
elements $\langle \chi_{k_{z}'}\mid v({\bf Q}=0,z)\mid \chi_{k_{z}}\rangle$
and $\langle \chi_{k_{z}'}\mid - (\partial/\partial z) v({\bf Q}=0,z)\mid
\chi_{k_{z}}\rangle$ which determine the parallel and perpendicular
components of expression (\ref{eq:F}), respectively. The minima occurring 
in these plots are the consequence
of the competition between the contributions coming from the repulsive and
attractive components of $v({\bf Q}=0,z)$.

As is clear from Eq. (\ref{eq:V}), the calculations of the full 
interaction matrix elements 
also require the dynamical
displacements ${\bf u_{l}}$ of atoms in the adlayer. For Xe adsorption on
Cu(001) the overlayer structure is incommensurate with the underlying
substrate which necessitates an approximate but consistent 
description of the adlayer 
vibrational properties. Following the experimental evidence 
indicating a defect free and well 
structured monolayer (c.f. Fig. \ref{nxefg2}) we
assume as in Sec. \ref{sec:expresul} a planar "floating adlayer" 
with perfect hexagonal 
symmetry in which the phonon dynamics 
takes into account only the averaged Xe-substrate potential
perpendicular to the surface as well as 
inter-adsorbate interactions. In this case the polarization
vectors and eigenfrequencies are solutions of a three-dimensional dynamical
matrix \cite{Woell}. Hence, the adsorbate displacements can be expanded
in terms of normal 
phonon modes of the adlayer in terms of phonon creation
($a^{\dag}_{{\bf Q},j}$) and annihilation ($a_{{\bf Q},j}$) operators:

\bq
{\bf u_{l}}=
\sum_{{\bf Q},j}
\left(\frac{\hbar}{2M_{a}N_{a}\omega_{{\bf Q},j}}\right)^{1/2}
e^{i{\bf Q\brho_{l}}}{\bf e}({\bf Q},j)(a_{{\bf Q},j}+a^{\dag}_{{\bf -Q},j}),
\label{eq:phonexp}
\eq
which also defines ${\bf u}({\bf Q},j)=
(\hbar/2M_{a}N_{a}\omega_{{\bf Q},j})^{1/2} {\bf e}({\bf Q},j)$ in Eq. 
(\ref{eq:V}). Here
${\bf \brho_{l}}$ is the parallel component of the radius vector of the
${\bf l}$-th adsorbate in equilibrium, and $M_{a}$ and $N_{a}$ are the
mass and the number of Xe atoms in the adlayer, respectively. In the 
floating adlayer model 
the index
$j$ ranges over the S- (shear vertical), L- (longitudinal horizontal)
and SH- (shear horizontal) adlayer modes.
By construction the floating adlayer phonon modes are  localized in the
adlayer and the polarizations of L- and SH-modes are strictly in the 
surface plane. 
The corresponding dispersion curves calculated with the adjusted force 
constants quoted in 
Sec. \ref{sec:expresul} are shown in Fig. \ref{nxefg8}. 
Consequently, the symmetry selection 
rules for probability of phonon excitation in a floating adlayer 
are more stringent than in 
the case of a dynamical matrix
description which includes the substrate dynamics in some way. 
This is reflected in the 
calculated probability of excitation of an SH-phonon for 
in-the-sagittal plane inelastic 
scattering within the first SBZ
of the superstructure. This probability turns out to be much 
smaller than for the L-mode 
for all azimuths, and is exactly zero along the two high symmetry directions 
$\Gamma \bar{K}_{Xe}$ and $\Gamma \bar{M}_{Xe}$ of the superstructure. 
These effects will be 
further discussed below in conjunction with the effects which the 
scattering potential imparts 
to the scattering intensities and in Sec. \ref{sec:exptheor} where we discuss a comparison between experimental and theoretical results.

To calculate the required eigenfrequencies and the corresponding 
polarization vectors of the 
full dynamical matrix corresponding to the commensurate Xe/Cu(111) 
system we have used the 
force constants
given in the previous section. The results 
for the dispersion curves are shown in Fig. \ref{nxefg9}.a.
Here we can also trace how each phonon mode of the
composite system is localized at the surface (i.e. within the adlayer) and
how typical surface modes may get delocalized for certain values of the
wave vector. 
A measure of the adlayer localization of the S-, L- and SH-modes near the
center of the first BZ of the superstructure is shown in Fig. \ref{nxefg9}.b.
Further important information concerning the ellipticity of polarization
(perpendicular versus longitudinal) of the L-mode in the same region 
is illustrated in 
Fig. \ref{nxefg9}.c.

The calculation of the force matrix elements, Eq.  (\ref{eq:F}), can
be performed analytically \cite{Devonshire} using
the wave functions of Eq. (\ref{eq:eigenstates}) and the two
dimensional Fourier
transform of the pair interaction, Eq. (\ref{eq:v}) \cite{Siber}. 
With the aid of these
matrix elements and using Eqs. (\ref{eq:calV}) and
(\ref{eq:Kronecker}), we can obtain first order or distorted wave Born
approximation (DWBA) transition probabilities for He atoms in Eq. 
(\ref{eq:WEBA}). 
Multiplying them by the corresponding Bose distributions we obtain 
the DWBA inelastic 
state-to-state reflection coefficients for one phonon scattering 
\cite{Cellirev}. Carrying 
out the summation over $k'_{z}$ by making use of Eq. 
(\ref{eq:Kronecker}), fixing 
$\theta_{SD}=\theta_{i}+\theta_{f}$, $E_{i}$ and $T_{s}$, 
and varying $\theta_{i}$ we obtain 
the $Q$-dependent scattering intensities which can be directly 
related to the one phonon 
intensities in the TOF spectra (c.f. next section). Here we shall 
only illustrate some of 
their general and most interesting features.

Figure \ref{nxefg10} shows a plot of the DWBA scattering intensities 
for emission (energy loss)
or absorption (energy gain) of a single dispersionless S-phonon in HAS 
from the
"floating" Xe adlayer on Cu(001) surface. The calculated intensities 
reveal a relatively simple 
and expected structure
as a function of the exchanged phonon momentum which is mainly due to
simple properties of S-phonons  characteristic of the "floating" adlayer model
(absence of dispersion and the polarization vector localized in and
perpendicular to the adlayer). On the other hand, due to the more 
complicated model of 
vibrational dynamics of the
commensurate
Xe monolayer adsorbed on Cu(111) surface the analogous intensities 
shown in Fig.
\ref{nxefg10}.b exhibit a more complicated structure despite the 
similarity in the 
corresponding
potential parameters (for the magnitude of the latter see next section).
Here the S-phonon polarization vector becomes delocalized over the first few
layers of the Xe/Cu slab for the values of $Q$ at which the S-phonon
dispersion curve meets the dispersion curves of substrate phonons
(c.f. Fig. \ref{nxefg9}.a). This makes the coupling of the
He atom to S-phonons weaker in this region of the $Q$-space which is then
reflected in the occurrence of pronounced minima in the scattering
intensities.

However, for nondispersive S-phonons the intensities of the measured single
quantum loss or gain peaks acquire additional weight due to the
multi-quantum interference between the emission of  $n \pm 1$ and 
annihilation of
$n$ nondispersive phonons of different wave vectors 
\cite{Kasai,MansonEinst}. These additional 
contributions are
automatically included in the EBA with account of recoil effects,
as demonstrated in Ref. \cite{1moment}.
The difference between the single S-phonon loss and gain scattering
intensities obtained in first order DWBA theory and in the
EBA (which takes into account such interference processes) can also be
visualized in Figs. \ref{nxefg10}.a and \ref{nxefg10}.b for 
incommensurate and commensurate Xe
adlayers, respectively. 
It is seen from the figures that for given projectile incoming
energies these differences may be already quite substantial. In particular,
the noticeably larger EBA intensities for larger wave vectors
are solely due to multi-quantum interference processes.
This necessitates the use of the EBA in calculating the scattering
intensities for comparison with the experimental data. On the other hand,
we have found that the same type of renormalization of the scattering
intensities of dispersive L-modes by emission and absorption of
nondispersive S-modes gives a negligible effect.

The differences in magnitude between the projectile transition 
probabilities in one phonon loss 
and gain processes can be best illustrated by repeating the above 
calculation for the 
scattering intensities but without multiplying the phonon loss and 
gain transition 
probabilities by the corresponding Bose distributions. The results 
of such a calculation for 
an S-mode emission and absorption are depicted in the insets in 
Figs. \ref{nxefg10}.a and 
\ref{nxefg10}.b. These differences arise
from and give a measure of the quantum recoil effects and are more pronounced
at lower projectile incoming energies (c.f. Fig. (6) in Ref.\cite{Xe}). The
trend that the results for loss processes give smaller values than 
for the gain processes 
reflects the
fact that the phase space or density of states for transitions of the
projectile to a state with lower energy (case of phonon emission) is smaller
than to a state with higher energy (case of phonon absorption).

Figure \ref{nxefg11}.a shows first order or DWBA intensities for HAS from
L-phonons in the incommensurate adlayer of Xe on Cu(001)
surface for finite substrate temperature $T_{s}$ which determines 
the temperature of the 
Bose-Einstein distributions for adlayer phonons in expression 
(\ref{eq:WEBA}). In the present 
"floating" adlayer model for the dynamical matrix the L-mode frequency follows
acoustic dispersion
$\omega_{{\bf Q},L}\propto c_{L}Q$ for small $Q$ (c.f. Fig.
\ref{nxefg8}). The corresponding
polarization vector ${\bf e}({\bf Q},L)$ lies strictly in the adlayer
plane and remains dominantly parallel to
${\bf Q}$ even outside the high symmetry directions of the 
BZ of the superstructure.
Consequently, the L-phonon DWBA intensity for small $Q$ 
(for which $\omega_{{\bf Q},L}\propto Q$) and finite $T_{s}$ 
becomes proportional to the factor 
$\mid{\bf Q}\cdot{\bf e}({\bf Q},L)\mid^2 kT_{s}/\omega^{2}_{{\bf
Q},L}$ which 
saturates at a finite value
for $Q\rightarrow 0$. As a result of that and the properties of the 
matrix elements depicted 
in Fig. \ref{nxefg7}.a the L-phonon scattering intensity in the case 
of the incommensurate 
monolayer exhibits a maximum near the zone center. The minima in the 
intensity occur for 
those values of $Q$ at which the on-the-energy-shell counterparts of 
these matrix elements 
go through a minimum or zero. All these features clearly
manifest themselves in the $Q$-dependence of the L-phonon
scattering intensity shown in Fig. \ref{nxefg11}.a.
In contrast to the L-mode behavior, the polarization of the SH-mode 
remains dominantly 
perpendicular to ${\bf Q}$ also
outside the high symmetry directions of the first SBZ of the 
superstructure \cite{Gumhalter}.
Therefore for ideally structured "floating" adlayers it will 
always give a much smaller 
contribution to the intensity
of the one
phonon processes for HAS in the sagittal plane and ${\bf Q}$ 
restricted to the first SBZ 
(c.f. next section).

Figure \ref{nxefg11}.b shows the single L-phonon HAS intensity 
for Xe/Cu(111) system as a 
function of the
phonon wave vector
on the same scale as in Fig. \ref{nxefg11}.a.  Apart from the 
trends leading to zero
scattering intensities for some isolated Q-values due to the 
behavior of the off-shell matrix 
elements
shown in Fig. \ref{nxefg7}.a, which are common to both 
incommensurate and
commensurate Xe layers, some basic differences with respect 
to the Xe/Cu(001) system can be 
observed.
The commensurability of the adlayer with the substrate gives 
rise to
non-vanishing Xe-Cu shear stress force constants entering the full dynamical
matrix of the Xe/Cu slab, thereby producing two important effects
regarding the phonon dynamics. First, it causes the appearance of a zone
center phonon gap in the dispersion curves of L-phonons  and 
hence the intensity factor
$\mid{\bf Q}\cdot{\bf e}({\bf Q},L)\mid^{2}kT_{s}/\omega^{2}_{{\bf
Q},L} \propto Q^{2}$ for 
small $Q$ because $\omega_{{\bf Q}=0,L}\neq 0$.
Second, the polarization vector of the L-phonons is no longer
constrained to the surface plane but for the values of $Q$ at
which the L-mode and substrate modes are degenerate it also 
acquires a component
in the direction perpendicular to the surface and its localization 
to the
adlayer is reduced (c.f. Figs. \ref{nxefg9}.b and \ref{nxefg9}.c). 
For a completely
in-surface-plane polarization of the L-mode and finite $T_{s}$ the 
first effect causes a
drop in the scattering intensity which is quadratic in $Q$ towards 
the zone center.
The beginning of this trend is clearly visible in Fig. \ref{nxefg11}.b.
However, by increasing $Q$ the second effect begins to play a role and, since
the coupling to perpendicular vibrations is much stronger than to the
parallel ones, the L-phonon scattering intensity rises rapidly up and reaches
a maximum in two spikes near the zone center. Hence, the interplay 
between the parallel and 
perpendicular polarizations, or the ellipticity of the L-mode 
polarization in the
commensurate Xe/Cu(111) system, introduces fast variations of the
scattering intensity as a function of the exchanged phonon momentum.
Regarding the role of the SH-modes of the slab on the HAS intensities, we find
that the situation here is completely analogous to the case of a floating
Xe adlayer, i.e. within the present model description their coupling 
to He atom is generally 
negligible for in-sagittal-plane collision geometry and ${\bf Q}$ 
lying in the first SBZ, and 
is strictly zero along the high
symmetry directions.
However, this conclusion does not apply to the magnitude of the Debye-Waller
factor because the Debye-Waller exponent is obtained by carrying out
the momentum transfer summations over the entire first SBZ and also
beyond if the coupling matrix elements are strong there
(c.f. Figs. \ref{nxefg10} and \ref{nxefg11}). In fact, for the 
matrix elements of appreciable 
magnitude in the second SBZ the SH-mode can produce even larger
contributions to the DW exponent than the L-mode  due to its lower 
excitation frequency. This 
situation is illustrated in Fig. \ref{nxefg12}.

An important point to be observed in connection with Figs. 
\ref{nxefg10} and \ref{nxefg11} is a
relatively large difference between the one phonon loss and 
gain scattering
probabilities at lower projectile incident energies. This is 
due to an
interplay between the projectile recoil and temperature effects. 
The amplitude for a
transition of the projectile to a state with higher energy  is 
larger than
to a state with lower energy (c.f. insets in Figs. \ref{nxefg10}.a 
and \ref{nxefg10}.b) because 
of the larger phase space for the states of higher energy. 
Thus, for the same incoming energy the matrix elements 
$\mid {\cal V}(-)\mid^{2}$ for phonon 
absorption in expression (\ref{eq:WEBA}) for the scattering 
function will be generally larger 
than $\mid{\cal V}(+)\mid^{2}$ which describe phonon emission 
(c.f. also Fig. 6 in Ref. 
\cite{Xe}). 
On the other hand, the temperature effects entering expression 
(\ref{eq:WEBA}) through the 
corresponding Bose-Einstein distribution factors $\bar{n}$ and 
$(\bar{n}+1)$ for phonon 
absorption and emission,
respectively, act just in the opposite direction because 
$\bar{n}<(\bar{n}+1)$ . Hence, the 
total scattering intensity for phonon emission or absorption 
calculated from the scattering 
function (\ref{eq:WEBA}) depends on the trade
off between these two effects.

\section{Comparison of theoretical and experimental results}
\label{sec:exptheor}

A real test of the assumptions underlying the present model
interpretation of HAS from Xe monolayers on Cu(001) and Cu(111) surfaces
should come through a comparison of relative excitation intensities 
of the various
modes in the experimental and calculated scattering spectra. This is equally
relevant for the single and multiphonon scattering regimes.

In this section we first calculate the relative intensities of the 
various adlayer modes in
the single phonon scattering regime using the approaches described 
in the preceding sections. 
The parameters characterizing
the He-Xe potential in expression (\ref{eq:v}) are given by:

$D=6.60$ meV, $\alpha^{-1}=0.8202$ \AA, and $z_{0}=3.49$ \AA,
\\
for the commensurate system Xe/Cu(111), and

$D=6.40$ meV, $\alpha^{-1}=1.032$ \AA, and $z_{0}=3.6$ \AA,
\\
for the incommensurate system Xe/Cu(001).
These are not the parameters obtained from pairwise summation of pure 
He-Xe gas-phase potentials
\cite{Kleine} which yields:

$D_{gas}=7.2$ meV, $\alpha_{gas}^{-1}=0.77$ \AA, and $z_{0,gas}=3.51$ \AA,\\
 but slightly modified ones to produce a softer He-surface potential. 
The present set of 
parameters for the Xe/Cu(111) system also differs from the one
quoted in Ref. 
\cite{BraunPRL} which is a result of additional consistency requirement
imposed to obtain a 
unified set describing the TOF spectral intensities equally well in the
single and multiphonon 
scattering regimes.
Such a necessity for modification of the sum
of
gas-phase pair potentials is not uncommon in HAS studies 
(c.f. Refs. \cite{Tommasini,modifiedpot}) and here it is
also necessary to consistently reproduce the relative peak intensities 
in the measured TOF 
spectra.

A comparison of the experimental and calculated spectral intensities of
S- and L-modes in the 
single phonon scattering regime of HAS from Xe/Cu(111) is shown in Fig.
\ref{nxefg13}. Here we note that our calculations always yield
smaller intensities for the elastic peaks compared with the experiment
as the latter also includes contributions from diffuse elastic
scattering from defects not accounted for by the present model. Hence,
this missing component has been added
to the elastic peak intensity because in combination
with the finite peak width it can also contribute to the background 
intensity of the L-peaks.
With this proviso a good agreement between experimental and
theoretical results is achieved which illustrates the consistency of 
the present interpretation 
of the inelastic  peaks in the TOF spectra.

Since in the floating adlayer model applied to the Xe/Cu(001) system the
interference between the adlayer modes and the substrate
Rayleigh wave (RW) cannot be obtained, we have selected the TOF spectra
in which the peaks
assigned to S- and L-modes could be maximally separated from those of 
the RW and then 
calculated only the adlayer mode intensities in the EBA by neglecting 
the S-mode frequency 
shifts and delocalization occurring at avoided crossing with the 
dispersion curve of the RW. 
Along the
direction of measurement the polarization of the SH mode is not 
strictly perpendicular
to its wave vector, and this in principle could give
rise also to SH-mode-induced structures in the TOF spectra.
However, in this direction the present theoretical analysis gives 
the SH-mode intensity of the 
order of only five percent relative to the contribution from the L-mode. Only the introduction of coupling of the SH mode to the modes of the underlying substrate could induce a breakdown of symmetry leading to a removal of such stringent selection rules and thereby to a larger SH-mode excitation probability.
The calculated relative intensities of the S- and L-modes for the Xe/Cu(001)
system in the single phonon scattering regime, but with inclusion of
multi-quantum interference of nondispersive S-modes, are compared with the
experimental data  in Fig. \ref{nxefg14}. Given all the approximations 
used in the calculation, 
it is seen that the
model reproduces the relative TOF intensities quite satisfactory. The 
only exception occurs on 
the energy loss side in the lower panel of Fig. \ref{nxefg14} where 
the afore mentioned avoided 
crossing between the substrate RW (energy loss at 3.05 meV) and the 
adlayer S-mode takes place,  with the effect of S-mode frequency shift 
and intensity enhancement. 
In Fig. \ref{nxefg14} we also demonstrate how the use of the unsoftened
force constants derived 
from the three-dimensional Xe-Xe gas phase potential \cite{HFD-B2} 
produces the position and 
intensity of L-peaks for which the disagreement with the experimental 
data is evident. On the 
other hand, the SH-peaks calculated for the present scattering geometry
by using the same gas 
phase potentials are of negligible relative intensity to be
experimentally observable although 
their frequency may coincide with that of the measured acoustic mode.   
Hence, with the present mode assignments and the corresponding 
model dispersion relations 
based on the softened radial intralayer force constants we can 
consistently describe the 
HAS-TOF intensities for the Xe/Cu(001) system as well.

As the coupling of He atoms to S-modes is much stronger than to the 
L-modes, as illustrated in 
Figs. \ref{nxefg10} and \ref{nxefg11}, the multiphonon
scattering spectra will be dominated by a series of multi-quantum S-peaks.
All other dispersive modes may only add weak structures on top of this
basic one. Eventually, these structures will turn into a broad Gaussian-like
background (cf. Eq. (73) in Ref. \cite{HAS} and Eq. (9) and Fig. 3 in Ref.
\cite{comment}) in the limit of high incident projectile energies.
The multiphonon scattering spectra from Xe/Cu(111) and Xe/Cu(001) adlayers
have been studied in detail in Ref. \cite{Xe} and for the sake of
completeness we here briefly illustrate their  features in Figs. 
\ref{nxefg15}.a and
\ref{nxefg15}.b, respectively, by consistently employing the potentials from
the corresponding single phonon calculations.
Figure \ref{nxefg15}.b is interesting in that it illustrates the behavior of
the experimental scattering spectra of He$\rightarrow$Xe/Cu(001) collisions
in the single phonon scattering regime regarding the modes  which weakly
couple to He atoms. This is shown by the appearance of a Rayleigh
wave-induced hump and an L-mode-induced shoulder near the elastic 
line. However, 
this scattering regime simultaneously
appears to be a
multiphonon one for the S-modes whose
coupling to the scattering He atoms is much stronger.
Concerning the agreement between experimental and theoretical results we
observe that although the multiphonon
TOF spectra of the incommensurate Xe/Cu(001) system can be
relatively well reproduced, except for the elastic line which was not 
corrected for diffuse scattering
contribution in the theoretical plot, the agreement for the commensurate
Xe/Cu(111) system is 
better. Interestingly enough, in the latter case (Fig. \ref{nxefg15}.a)
the diffuse elastic
scattering correction for the elastic line was unnecessary due to the 
true multiphonon 
character of the spectrum at
this higher incident energy. 
 This spectrum can be viewed as a convolution of a series of well defined
equidistant peaks, signifying the uncorrelated multiple
emission and absorption of nondispersive S-phonons and not the overtones 
(for overtone frequencies see next section),
with a broad background arising from the 
multiple excitation of L- and SH-phonons (whose polarizations in 
this regime are no 
more constrained by the symmetry selection rules), and also from the substrate surface projected modes which may couple to the scattered He atoms. As implied in 
expression (\ref{eq:WEBA}), 
the phonon absorption processes of any kind can take place only for $T_{s}>0$.

It is also noteworthy that the resulting multi-quantum S-mode intensities in
the spectra shown in Figs. \ref{nxefg15}.a and \ref{nxefg15}.b do 
not generally follow the 
Poisson distribution.
This arises as a consequence of the projectile recoil effects and the
dependence of the magnitude of each peak maximum on the exchanged parallel
momentum, which both act so as to prevent the appearance of a simple
Poissonian structure. These features were discussed in detail in Refs.
\cite{Xe,1moment,Mats}.

\section{Corrugation of X\lowercase{e}-C\lowercase{u}(111) potential 
energy surface}
\label{sec:corr}

The value of the zone center gap for the longitudinal phonon mode in 
the case of Xe adsorbed on
Cu(111) surface also provides information on  the Xe/Cu(111) potential 
energy surface.
Since, in the past, precise information on the lateral corrugation for 
noble gases adsorbed on
metal surfaces has been missing, an approach
based on the work by Steele \cite{Steele} has frequently been used in 
applications 
\cite{Robbins} in which the periodic adsorbate-substrate 
potential $V_{s}({\bf r})$ is obtained by a summation of Lennard-Jones 
pair potentials 
$V({\bf r})=4\epsilon[(\sigma/r)^{12}-(\sigma/r)^{6}]$ describing 
interaction between noble
gas atom and substrate atoms. The resulting sum can be expressed in the
form

\bq
V_{s}(\brho,z)=\epsilon [V_{0}(z)+f\, V_{1}(z)
\sum_{i=1}^{3}\cos({\bf G}_{i}\cdot \brho)].
\label{eq:corrpot}
\eq
where now $\brho$ and $z$ denote the adsorbate parallel and 
perpendicular coordinate, 
respectively, the latter being measured from the first layer 
of substrate atoms. The symbols 
$\pm{\bf G}_{i}$, $i=1$ to 3, denote the six shortest reciprocal
lattice vectors, and the definition of $V_{0}$ and $V_{1}$ is given in Ref. \cite{Steele}. The original expression (\ref{eq:corrpot}) was derived in Ref. \cite{Steele} for the case $f=1$ and the extra
factor $f$ which may differ from unity has been added following Ref. 
\cite{Robbins} to adjust the corrugation
without changing the adsorption energy.
Since this type of the potential has been used in molecular dynamics simulations of sliding monolayers we can estimate its relevance to the present problem even without invoking the precise information on the adsorption sites. Thus, setting $\sigma$ to
3.487 \AA, the arithmetic mean of  $\sigma_{Xe-Xe}$ (describing the Xe-Xe
interaction), the Cu nearest neighbor distance
\cite{Crow} and putting $\epsilon=19$ meV, yields a
Xe-surface potential with a total depth of around 200 meV, a reasonable
value \cite{Xe:pot:theory}.
For the interaction between noble gas atoms
and metal surfaces  with highly delocalized conduction electrons
the
parameter $f$ has previously been reduced to values of the order 0.1 
to account
for small corrugations observed experimentally, and values as small 
as 0.03 have been proposed
for the case of Kr on Au(111) \cite{Robbins}.  

Although we have not used the parametrized expression (\ref{eq:corrpot}) in our dynamical matrix analysis of Xe adlayer vibrations, 
we would like to point out that for the potential parameters appropriate to
Xe on Cu(111) (see above) a value of $f=0.1$ yields a zone center phonon gap
energy  of 0.25 meV,
significantly  smaller than the value reported here.
In addition, using this potential energy surface (PES) the computed S-phonon energy  amounts to 
4.55 meV (i.e. much larger than 
the present experimental value $\hbar\omega_{S}=2.62$ meV), and the anharmonic shift of the fifth overtone to $\sim$0.5 meV. The same conclusion can be also derived by using the Xe-Cu(111) potentials parametrized in Refs. \cite{Xe:pot:theory,Xe:pot:exp}. 
To get the correct curvature of the PES at the adsorption site one should either change by a substantial amount the depth of the potential (\ref{eq:corrpot}) in the normal direction (governed by $\epsilon$), or modify its width which is controlled by $\sigma$, or both. Drastic changes in the potential depth leading to the experimental vibrational energies are physically implausible in view of the correct fitting of the desorption energies. 
On the other hand, the correct curvature can be achieved by a modest, 10\% softening of the repulsive component of the potential. This is consistent with the softening of the L-mode force constants discussed in Sec. \ref{sec:expresul} and the modification (also in the direction of softening) of the He-Xe adlayer potential noted in Sec. \ref{sec:exptheor}. All three features, which derive from independent analyses of the experimental data, unambiguously point towards the effect of softening of pair interactions involving Xe atoms in adsorbed monolayer phase.

In the previous report of the present results for Xe/Cu(111) 
\cite{BraunPRL}  an
incorrect reference was made to the work of Cieplak, Smith and Robbins
\cite{Robbins} in which the results of molecular dynamics simulations for
Kr adsorbed on Au(111), i.e. not Xe on Ag(111),
were reported. Considering the
analysis presented in the last paragraph, we believe that the PES 
used in that work 
\cite{Robbins} may be too weakly corrugated when the factor $f$ is 
reduced to below
the value of 0.1. Since the molecular dynamics results on  friction 
of sliding noble gas 
adlayers
show a pronounced dependence  on the strength of lateral corrugation 
\cite{Robbins,Persson}
it would be highly desirable to experimentally determine the zone center
gap of the longitudinally polarized  phonon modes of 
Kr and Xe adsorbed on
Ag(111) or Au(111).
Unfortunately, however, such measurements 
are hampered by the fact that on these substrates
both  noble gases form incommensurate
overlayers in which the zone center phonon gap is zero because of
the translational invariance.

\section{Summary and Conclusions}
\label{sec:concl}

In this work we have carried out a comprehensive comparative experimental
and theoretical study of low energy dynamics of monolayers Xe on Cu(111)
and Cu(001) surfaces by utilizing the HAS-TOF spectroscopy and a 
recently developed 
quantum theory of inelastic HAS from surfaces which treats single 
and multiphonon scattering 
processes on an equivalent footing. The inelastic HAS data were
obtained for a wide range of initial scattering conditions ($E_{i},
\theta_{i}, T_{s}$) spanning the single and multiphonon scattering
regimes.
The data have been carefully analyzed by combining the
dynamical matrix approach for description of the adlayer vibrational
dynamics with the developed scattering theory.

For both substrates the angular distributions (diffraction spectra) of
scattered He atoms signified hexagonally ordered Xe monolayers,
the commensurate ${\rm (\sqrt{3}\times\sqrt{3})R30^{0}}$ adlayer in 
the case of
the Cu(111) surface \cite{Chestersetal,Diehl} and the
incommensurate one in the case of the Cu(001)
surface\cite{Chesters,Glachant}.
The measured phonon dispersion relations deduced from the TOF spectra for
the two types of monolayers exhibited great similarities in the case of the
nondispersive optical-like S-phonons localized in the adlayer. However, a
striking difference occurred in the case of the very soft dispersive phonon
branch labeled "L". In the incommensurate phase it is genuinely
acoustic-like whereas in the commensurate phase it exhibits a frequency
band gap at
the zone center.

The tentative mode assignment was first made by employing the symmetry 
selection rules 
applicable to one-phonon excitation processes in HAS in the sagittal 
plane in combination 
with the dynamical matrix description of the adlayer vibrational modes.
Since in the case of the Xe/Cu(001) system no signature of a possible high
order commensurate phase has been found,
the dynamical matrix treatment of the two adlayers was markedly different.
On Cu(001) it was only possible to treat incommensurate Xe adlayer as 
an ideally structured 
hexagonal monolayer floating on a
structureless substrate surface with adlayer modes completely decoupled
from those
of the substrate. 
On the other hand, for the monolayer of Xe on Cu(111) it was
possible to construct a full dynamical matrix of the vibrationally coupled
Xe/Cu(111) system. Despite these differences it was possible to establish a
unified model interpretation of the observed modes in both systems in 
terms of a
consistent set of adjusted Xe-Xe and Xe-Cu force constants. In this picture
the S-mode is dominantly adlayer localized and $FT_{z}$ or vertically 
polarized.
In order to reconcile the symmetry selection rules for excitation of 
in-plane-polarized 
adlayer-induced modes in
HAS from ideally
structured adlayers with the dynamical matrix description of the modes,
we had to assign a 
dominantly longitudinal polarization
to the "L"-mode (L-phonon), with a possibility of a
relatively strong vertical or $z$-admixture (up to $\sim 15\%$) near the
avoided crossings with other surface projected modes.
By consistently carrying out this procedure we found that the radial
Xe-Xe force constants 
determining the dispersion
of the thus assigned longitudinal modes turned out much softer than 
by directly applying the 
sophisticated HFD-B2 Xe-Xe gas phase
pair potential\cite{HFD-B2}. Such a modification of the interadsorbate 
interactions still 
awaits its interpretation
through a detailed
calculation of the electronic structure of Xe adlayers on Cu(111) and Cu(001).
The magnitude of the band gap in the dispersion of L-phonons in the
commensurate Xe/Cu(111) phase was used to obtain information on the
corrugation of the Xe atom-substrate potential in this system\cite{BraunPRL}.

The assignment of the observed acoustic modes to shear horizontal or 
SH-modes in the first SBZ, 
which were the subject of recent controversy\cite{Bruch}, was ruled 
out for both systems first 
by invoking the symmetry selection rules (c.f. Sec. \ref{sec:expresul})
for the excitation of 
the various in-plane modes in HAS. Following these arguments, in the 
present model in which 
the ideal structure of Xe adlayers is an essential ingredient, the 
SH-mode can give only a 
negligible contribution to the scattering intensity relative to the 
longitudinal mode.

These interpretations were then corroborated by detailed theoretical 
analyses of the
scattering intensities in the HAS-TOF spectra. We have first analyzed and
demonstrated the difference between the interaction matrix elements
coupling the scattering He atoms to the perpendicular and parallel to the
surface  vibrations of the adlayer for small parallel momentum transfer and
variable incoming projectile energy.
Combining this with the calculated properties of the polarization vectors
of the adlayer modes, we were able to obtain the one-phonon
HAS intensities of the adlayer modes and compare them with experiments. A
good and consistent agreement between experimental and model
theoretical results for the S- 
and L-modes 
was obtained in all aspects of the measured data for both types of 
Xe adlayers.

The theory was then extended to calculations of the scattering spectra
in the transition from the single to the truly multiphonon scattering
regime which is characterized by the value of the corresponding
Debye-Waller exponent larger than unity.
Interestingly enough, for the S-modes, which strongly couple to the 
scattered He
atoms, the multiphonon regime is already reached for the scattering
conditions which still favor the single L-phonon scattering. The 
interplay of these two 
types of couplings gives
rise to the characteristic spectral shapes of the multiphonon HAS TOF
spectra. A very good agreement between
the calculated results and the TOF data in this regime gives further 
support to
our earlier assignments, the consistency of the model
description of the low energy dynamics of the two distinct monolayer 
phases of Xe, 
and the adequacy of the present treatment of single and multiphonon He atom
scattering from these adlayers.

\acknowledgments

The work in Zagreb has been supported in part by the National Science
Foundation grant JF-133, and the work in Bochum in part by the German DFG
grant Wo~464/14-1.

\newpage

\begin{figure}
\caption{(a) Right panel: Structure of 
${\rm (\protect\sqrt{3}\times\protect\sqrt{3})R30^{\circ}}$ 
monolayer of Xe atoms (shaded circles) on (111) surface of Cu crystal, with two high symmetry directions (azimuths) 
in the substrate surface plane denoted. Left panel: 
Two dimensional Brillouin zones of the Cu(111) surface
(dashed lines) and of Xe adlayer (full lines). 
(b) He-atom angular distribution along the $[1 \bar{1} 0]$ azimuth of Cu(111) from Xe  
${\rm (\protect\sqrt{3}\times\protect\sqrt{3})R30^{\circ}}$ monolayer for incident wave vector $k_{i}=9.2$
 \AA$^{-1}$ ($E_{i}$=45~meV) and surface temperature 60~K. 
Peaks normalized to the height of the specular peak.
}
\label{nxefg1}
\end{figure}

\begin{figure}
\caption{He-atom angular distribution along the $[100]$
azimuth of the substrate from an incommensurate hexagonal monolayer of Xe
atoms adsorbed on Cu(001) for  $k_{i}=5.25$ \AA$^{-1}$ ($E_{i}=14.36$ meV)
and substrate temperature 52 K.}
\label{nxefg2}
\end{figure}

\begin{figure}
\caption{Series of measured HAS TOF spectra for a monolayer of
Xe on Cu(111) along the $[11\bar{2}]$ direction of the substrate
surface for three representative He atom incident energies.
$\theta_{i}$ and $\theta_{SD}$ denote the angle of incidence and the 
fixed total
scattering angle, respectively. Vertical scale measures relative peak heights in arbitrary units. 
Scattering parameters shown in the insets.}
\label{nxefg3}
\end{figure}

\begin{figure}
\caption{Series of measured HAS TOF spectra for a monolayer of Xe
on Cu(001) along the $[100]$ direction of Cu surface
for three representative He atom incident energies. Vertical scale in arbitrary units. Scattering
parameters shown in the insets.}
\label{nxefg4}
\end{figure}

\begin{figure}
\caption{Phonon dispersions for Xe/Cu(111) surface along the
$[11\bar{2}]$ direction relative to
the substrate as determined by HAS (full circles). The solid
line denotes the best fit achieved for the longitudinal (L) mode
in the Xe adlayer and the dashed-dotted line is the result for
the L-mode using the gas phase Xe-Xe potential. The theoretical
dispersion curve for the vertically polarized S-mode is marked by the 
long dashed line and of the Rayleigh phonon and the projected
bulk phonon edge of the Cu substrate by the full thin and dotted line, 
respectively.
For force constants see main text.}
\label{nxefg5}
\end{figure}

\begin{figure}
\caption{Phonon dispersions for Xe/Cu(001) surface along
the $[100]$ direction relative to the Cu substrate as
determined by HAS (full circles). The full curve
represents the best fit achieved for the longitudinal (L) mode,
the flat dashed curve shows a theoretical dispersion curve
for the S-mode, the dot-dashed line indicates the position
of the L-mode using the gas phase force constant. Dashed curve indicates the position of the Rayleigh
wave on the Cu(001) surface. For force constants see main text.}
\label{nxefg6}
\end{figure}

\begin{figure}
\caption{(a) Off-shell matrix elements of the potential  which couples the
projectile to in-plane surface vibrations (see text) for $Q=0$ and three
different values of the incoming perpendicular wave vector: $k_{zi}=3.5$
\AA (full line), $k_{zi}=7.0$ \AA (dashed line) and $k_{zi}=14.0
$ \AA
(dotted line), as function of scattered wave vector $k_{z}'$. Potential
parameters corresponding to Xe/Cu(111) system. (b) Same for projectile
coupling to perpendicular vibrations.}
\label{nxefg7}
\end{figure}

\begin{figure}
\caption{Calculated dispersion curves of S-, L- and SH-phonons (from top
to bottom) over 1/6 of the two dimensional first Brillouin zone, i.e. 
between two equivalent high symmetry directions, corresponding to a
floating Xe adlayer
on Cu(001). For force constants parametrization see main text.
The angle $\phi$ is measured relative to $\bar{\Gamma} \bar{M}_{Xe}$ 
direction of the Brillouin zone of the superstructure.}
\label{nxefg8}
\end{figure}

\begin{figure}
\caption{(a)  Dispersion curves for Xe/Cu(111) system along  the
$[11\bar{2}]$ direction calculated
using the dynamical matrix approach (see text) with
Xe atoms on-top of Cu atoms. Force constants parametrization
same as in Ref. \protect\cite{BraunPRL}.
Note the S-, L- and SH-mode dispersion curves detached from the bulk 
quasi-continuum.
(b) Surface localization of Xe induced S-, L- and SH-modes on the adlayer
expressed through the sum of the components of the respective
polarization vectors in the adlayer. Numbers below symbols 
$\protect\bigtriangleup$ denote frequencies (in meV) of the S-mode.
(c) Ellipticity (vertical vs. longitudinal polarization) of the L-mode 
in Xe adlayer on Cu(111) substrate as function of
phonon wave vector. Percentage above full squares denotes surface 
localization of the L-mode. }
\label{nxefg9}
\end{figure}

\begin{figure}
\caption{(a) Calculated DWBA and EBA scattering intensities for emitting one
S-phonon (lower and upper full curves, respectively, marked by $(+)$)
and absorbing one
S-phonon (lower and upper dashed curves, respectively, marked by $(-)$)
in HAS from
Xe/Cu(001) as function of exchanged phonon wave vector $Q$ for the scattering
conditions as denoted. For potential parameters see main text. Inset:
Corresponding projectile transition probabilities in one phonon loss
and gain processes
denoted by full and dashed lines, respectively.
(b)  Same for Xe/Cu(111) system.}
\label{nxefg10}
\end{figure}

\begin{figure}
\caption{(a) Calculated DWBA HAS intensity for L-phonon emission (full curve)
 and L-phonon absorption (dashed curve) in Xe/Cu(001) as function of
exchanged phonon momentum. For interpretation of the maxima and minima
see main text. (b) Same for Xe/Cu(111) system.}
\label{nxefg11}
\end{figure}

\begin{figure}
\caption{Plots of the various contributions to the Debye-Waller factor 
pertinent to He$\rightarrow$Xe/Cu(001) collision system as function of 
substrate temperature $T_{s}$ and for the scattering conditions as denoted. 
Dotted line: only S-phonons included; dashed line: S- and L-phonons
included; full line: S-, L- and SH-phonons included.
Note the logarithmic scale on the ordinate axis. Here the ${\bf Q}$
summation in Eq. (\protect\ref{eq:WEBA}) to obtain the DW exponent was 
carried out over the first and second surface Brillouin zone of the 
superstructure.}
\label{nxefg12}
\end{figure}

\begin{figure}
\caption{Comparison of experimental and calculated S- and L-mode
intensities in the single-phonon scattering spectra typical of
He$\rightarrow$Xe/Cu(111) collisions.}
\label{nxefg13}
\end{figure}

\begin{figure}
\caption{Comparison of experimental and calculated S- and L-mode
intensities in the single-phonon scattering spectra typical of He$\rightarrow$Xe/Cu(001) collisions. Expanded contours in the topmost 
panel show the positions and relative intensities of the L-mode
peaks calculated with the present softened Xe-Xe radial force constants 
(full curve) and unmodified gas phase
potential-derived \protect\cite{HFD-B2} force constants (diamonds). 
Similar differences are obtained for
other two spectra. SH-derived peaks not discernible on the present scale.}
\label{nxefg14}
\end{figure}

\begin{figure}
\caption{(a) Comparison of experimental and calculated EBA multiphonon
scattering spectra typical of He$\rightarrow$Xe/Cu(111) collisions  using
the same potentials and dispersion relations as in Fig. 
\protect\ref{nxefg13}.  
(b) Same for
He$\rightarrow$Xe/Cu(001) collisions using the potentials and
dispersion relations as in Fig. \protect\ref{nxefg14}.
}
\label{nxefg15}
\end{figure}


\begin{references}

\bibitem{Eigler} D.M. Eigler and E.K. Schweizer, Nature (London) 
{\bf 344},524(1990).

\bibitem{Rieder} G. Meyer, L. Bartels, S. Z\"{o}phel, E. Henze and 
K.-H. Rieder, Phys. Rev. Lett. {\bf 78},1512(1997).

\bibitem{PerssonTosatti} See articles in: {\em Physics of Sliding
Friction}, edited by B.N.J. Persson and E. Tosatti, Kluwer Academic 
Publishers, Dordrecht/Boston/London, 1996.    

\bibitem{WittePRL} G. Witte, K. Weiss, P. Jakob, J. Braun, K.L. Kostov
and Ch. W\"{o}ll, Phys. Rev. Lett.
{\bf 80},121(1998).

\bibitem{Kreuzer} See for example: W. Widdra, P. Trischberger, W.
Friess, D. Menzel, S.H. Payne and H.J. Kreuzer, Phys. Rev. 
{\bf B57},4111(1998); and references therein.   

\bibitem{Mason:Williams}
B.F. Mason and B.R. Williams, Phys. Rev. Lett. {\bf 46},1138(1981).

\bibitem{Sibener}
K. D. Gibson, S. J. Sibener, B. Hall, D. L. Mills, and J. E. Black,
J. Chem. Phys. {\bf 83},4256(1985).

\bibitem{Zeppenfeld}
P. Zeppenfeld, M. B\"uchel, R. David, G. Comsa, C. Ramseyer and C. Giradet,
Phys. Rev. {\bf B50},14667(1994).

\bibitem{Xe001} A.P. Graham, M.F. Bertino, F. Hofmann, J.P. Toennies
and Ch. W\"{o}ll,
J. Chem. Phys. {\bf 106},6194(1997).

\bibitem{BraunPRL} J. Braun, D. Fuhrmann, A. \v{S}iber, B. Gumhalter and Ch.
W\"{o}ll, Phys. Rev. Lett. {\bf 80},125(1998).

\bibitem{Gerlach} R. Gerlach, A.P. Graham, J.P. Toennies and H. Weiss, J. Cem. Phys. {\bf 109},5319(1998).

\bibitem{Bruch} L.W. Bruch, J. Chem. Phys. {\bf 107},4443(1997);
A.P. Graham, M.F. Bertino, F. Hofmann, J.P. Toennies, and Ch. W\"{o}ll,
J. Chem. Phys. {\bf 107},4445(1997);
L.W. Bruch, preprint (MPI f\"{u}r Str\"{o}mungsforschung,
G\"{o}ttingen, 1997).

\bibitem{HFD-B2} A.K. Dham, W.J. Meath, A.R. Allnatt, R.A. Aziz,
and M.J. Slaman, Chem. Phys. {\bf 142},173(1990).

\bibitem{Dove} M.T. Dove, {\em Introduction to Lattice Dynamics} 
(Cambridge University Press, Cambridge, 1993).

\bibitem{HAS} A. Bili\'{c} and B. Gumhalter, Phys. Rev.
{\bf B 52},12307(1995).

\bibitem{comment} B. Gumhalter and A. Bili\'{c}, Surf. Sci.
{\bf 370},47(1997).

\bibitem{BraunApp} B.J.~Hinch, A.~Lock, H.H.~Madden, J.P.~Toennies and
G.~Witte,
Phys. Rev. {\bf B 42},1547(1990).

\bibitem{Jupille} J. Jupille, J.-J. Erhardt, D. Fargues and A. Cassuto,
Faraday Discuss. Chem. Soc. {\bf 89},323(1990);
Vacuum {\bf 41}/No. 1-3,399(1990).

\bibitem{Xe} J. Braun, D. Fuhrmann, M. Bertino, A.P. Graham, J.P. Toennies,
Ch. W\"{o}ll, A. Bili\'{c} and B. Gumhalter, J. Chem. Phys.
{\bf 106},9922(1997).

\bibitem{Chestersetal} M.A. Chesters, M. Hussain and J. Pritchard, 
Surf. Sci. {\bf 35},161(1973).

\bibitem{Diehl} Th. Seyller, M. Caragiu, R.D. Diehl, P. Kaukasoina and 
M. Lindroos, Chem. Phys. Lett. {\bf 291},567(1998).

\bibitem{Chesters} M.A. Chesters and J. Pritchard, Surf. Sci.
{\bf 28},460(1971).

\bibitem{Glachant} A. Glachant and U. Bardi, Surf. Sci. {\bf 87},187(1979).

\bibitem{Woell} Ch. W\"{o}ll, Appl. Phys. {\bf A53},377(1991).

\bibitem{Harten} U. Harten, J.P. Toennies and Ch. W\"{o}ll, Faraday
Discuss. Chem. Soc. {\bf 80},137(1985).

\bibitem{Comsa}
B. Hall, D. L. Mills, P. Zeppenfeld, K. Kern, U. Becher and G. Comsa,
Phys. Rev. {\bf B40},6326(1989).

\bibitem{Xe:pot:theory}
A. Chizmeshya and E. Zaremba, Surf. Sci. {\bf 268},432(1992).

\bibitem{Xe:pot:exp}
G. Vidali, G. Ihm, H.-Y. Kim, M.W. Cole, Surf. Sci. Rep. {\bf 12},133(1991).

\bibitem{Cu001} G. Benedek, J. Ellis, N.S. Luo, A. Reichmuth, P.
Ruggerone and J.P.
Toennies  Phys. Rev. {\bf B48},4917(1993);
C.~Kaden, P.~Ruggerone, J.P.~Toennies, G.~Zhang and G.~Benedek,
Phys. Rev. {\bf B46},13509(1992).

\bibitem{deWette}
R. E. Allen, G. P. Alldredge, and F. W. de Wette, Phys. Rev. {\bf B4},
1648 (1971).

\bibitem{Cellirev} V. Celli in {\em Surface Phonons}, editors W. Kress and
F.W. de Wette (Springer, Berlin 1991), p.167.

\bibitem{Santoro} G. Santoro and V. Bortolani in {\em Inelastic Energy
Transfer in Interactions with Surfaces and Adsorbates}, edited by B.
Gumhalter, A.C. Levi and F. Flores (World Scientific, Singapore, 1993), p. 1.

\bibitem{Silvestri} A. Glebov, W. Silvestri, J.P. Toennies, G. Benedek
and J.G. Skofronick,
 Phys. Rev. {\bf B54},17866(1996).

\bibitem{Lang} N.D. Lang, Phys. Rev. Lett. {\bf 46},842(1981);
N.D. Lang and
A. R. Williams, Phys. Rev. {\bf B25},2940(1982).

\bibitem{Hall} B. Hall, D.L. Mills and J.E. Black, Phys. Rev. 
{\bf B 32},4932(1985).

\bibitem{Ellis} J. Ellis, J.P. Toennies and G. Witte,
J. Chem. Phys. {\bf 102},5059(1995).

\bibitem{Gumhalter} B. Gumhalter, unpublished.

\bibitem{surfstate} M. Wolf, E. Knoesel and T. Hertel, Phys. Rev. 
{\bf B54},R5295(1996).

\bibitem{BGL} K. Burke, B. Gumhalter and D.C. Langreth, Phys. Rev.
{\bf B47},12852(1993-I).

\bibitem{GBL} B. Gumhalter, K. Burke and D.C. Langreth, Surf. Rev. Lett.,
{\bf 1},133(1994).

\bibitem{VAS} J. Braun, D. Fuhrmann, J.P. Toennies, Ch. W\"{o}ll, A.
Bili\'{c} and B. Gumhalter, Surf. Sci. {\bf 368},232(1996).

\bibitem{Celli}
V.~Celli, D.~Himes, P.~Tran, J.P.~Toennies, Ch.~W\"{o}ll and G.~Zhang,
Phys.~Rev.~Lett. {\bf 66},3160(1991).

\bibitem{Manson}
J.R. Manson, V. Celli and D. Himes, Phys. Rev. {\bf B49},2782(1994); J.R.
Manson, Comput. Phys. Communications {\bf 80},145(1994).

\bibitem{BortoLevi} V. Bortolani and A.C. Levi, La Rivista del Nuovo
Cimento {\bf 9}/No. 11,1(1986).

\bibitem{BortolaniPRB}
V.~Bortolani, A.~Franchini, G.~Santoro, J.P.~Toennies, Ch.~W\"{o}ll and
G.~Zhang,
Phys.~Rev.~{\bf B40},3524(1989).

\bibitem{WitteRhodium}
G.~Witte, J.P.~Toennies  and  Ch.~W\"{o}ll,
Surf. Sci. {\bf 323},228(1995).

\bibitem{Brenig} W. Brenig, Z. Physik {\bf B 36},81(1979).

\bibitem{Meyer} H.-D. Meyer, Surf. Sci. {\bf 104},117(1981).

\bibitem{Brako} R. Brako, Surf. Sci. {\bf 123},439(1982).

\bibitem{Kern} K. Kern, P. Zeppenfeld, R. David and G. Comsa, J. Electron
Spec. Rel. Phen. {\bf 44},215(1987).

\bibitem{DWF} B. Gumhalter, Surf. Sci. {\bf 347},237(1996).

\bibitem{Evans-Mills} It has been shown that an approximate expression similar to (\protect\ref{eq:specEBA}) can be obtained also for the probability of electron scattering by optical surface phonons (see E. Evans and D.L. Mills, Phys. Rev. {\bf B7},853(1973)). The validity of such expression has been demonstrated in a passage from quantum-mechanical to quasiclassical description of the projectile motion  assuming long range projectile-phonon interaction. 
 

\bibitem{MansonArmand}(a): J.R. Manson and G. Armand, Surf. Sci.
{\bf 184},511(1987); (b): ibid. Surf. Sci. {\bf 195},513(1988).

\bibitem{group} V. Celli, G. Benedek, U. Harten, J.P. Toennies,
R.B. Doak
and V. Bortolani, Surf. Sci. {\bf 143},L376(1984).

\bibitem{TT} K.T. Tang and J.P. Toennies, Z. Phys. {\bf D1},91(1986).

\bibitem{Kleine} U. Kleinekath\"{o}fer, K.T. Tang, J.P. Toennies and C.L.
Yiu, Chem. Phys. Lett. {\bf 249},257(1996); U. Kleinekath\"{o}fer, Ph.D.
thesis, University G\"{o}ttingen, 1996; Max-Planck-Institut f\"{u}r
Str\"{o}mungsforschung Bericht 6/1996.

\bibitem{Gaspar} G. Ga\v{s}parovi\'{c}, B.Sci. thesis, University of Zagreb,
1997 (unpublished).

\bibitem{Siber} A. \v{S}iber and B. Gumhalter, Surf. Sci. {\bf 385},270(1997).

\bibitem{Hoinkes}  H. Hoinkes, H. Nahr and H. Wilsch, Surf. Sci. 33(1972)516,
ibid. 40(1973)129; G. Armand, J. Lapujoulade and Y. Lejay, Surf. Sci.
63(1977)143.

\bibitem{Tommasini} A. Franchini, G. Santoro, V. Bortolani, A. Bellman,
D. Cvetko, L. Floreano, A. Morgante, M. Peloi, F. Tommasini and T. Zambelli,
Surf. Rev. Lett. {\bf 1},67(1994).

\bibitem{Devonshire}  A.F. Devonshire, Proc. Roy. Soc. {\bf A 158},253(1937).

\bibitem{Kasai} H. Kasai and W. Brenig, Z. Phys. {\bf B59},429(1985).

\bibitem{MansonEinst} J.R. Manson, Phys. Rev. {\bf B37},6750(1988).

\bibitem{1moment} B. Gumhalter, A. \v{S}iber and J.P. Toennies, submitted
for publication.

\bibitem{modifiedpot} D. Eichenauer, U. Harten, J.P. Toennies and V. Celli,
J. Chem. Phys. {\bf 86},3693(1957).

\bibitem{Mats} C.M. Heden\"{a}s and M. Persson, Phys. Rev.
{\bf B45},11273(1992).

\bibitem{Steele} W.Steele, Surf. Sci. {\bf 36}, 317 (1973)

\bibitem{Robbins} M. Cieplak, E.D. Smith and M.O. Robbins, 
Science {\bf 265},1209 (1994)

\bibitem{Crow} A.D. Crowell and R.B. Steele, J. Chem. Phys. 
{\bf 34},1347 (1961).

\bibitem{Persson} B.N.J. Persson and A. Nitzan, Surf. Sci. 
{\bf 367},261(1996).

\end{references}
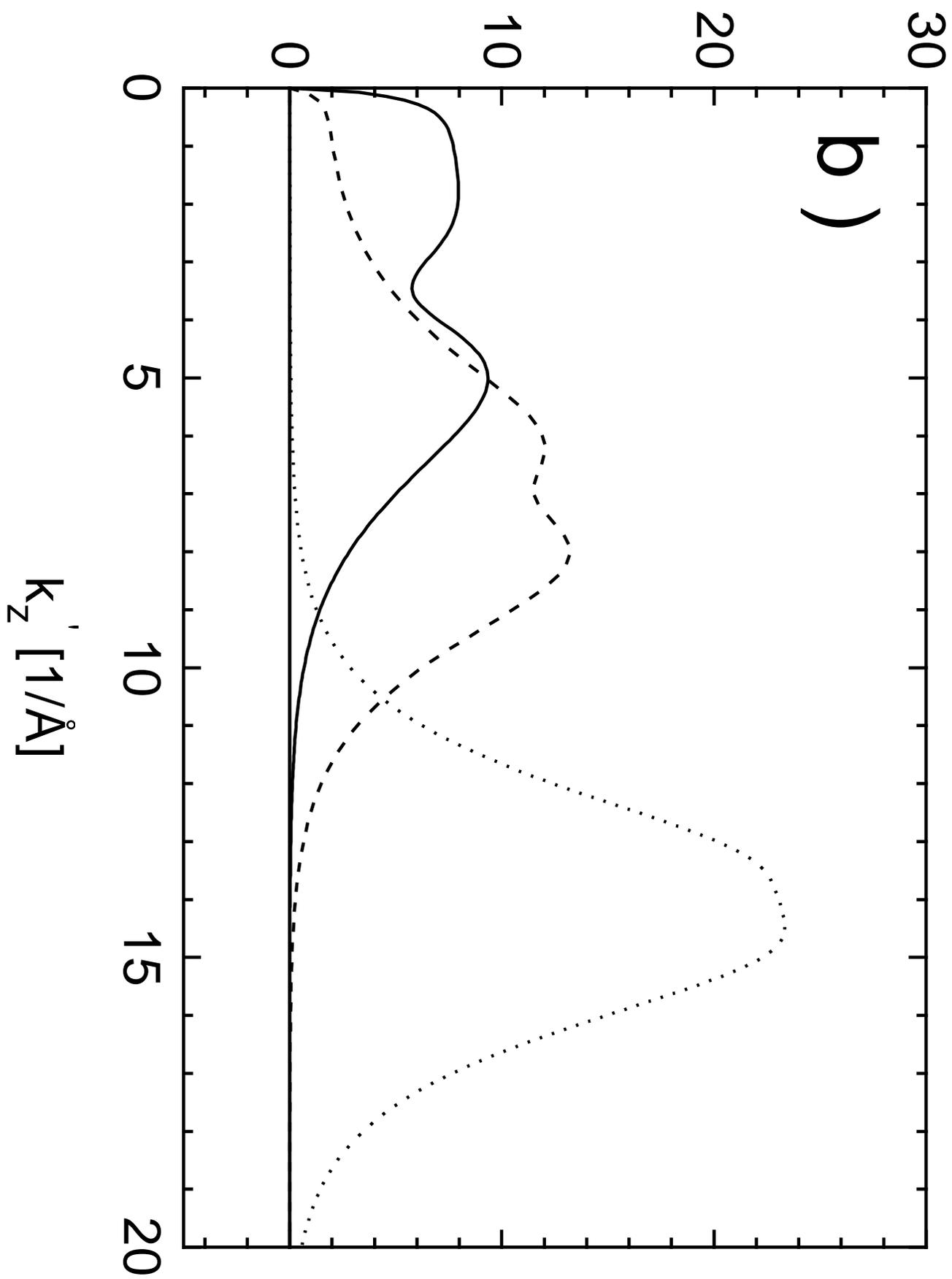
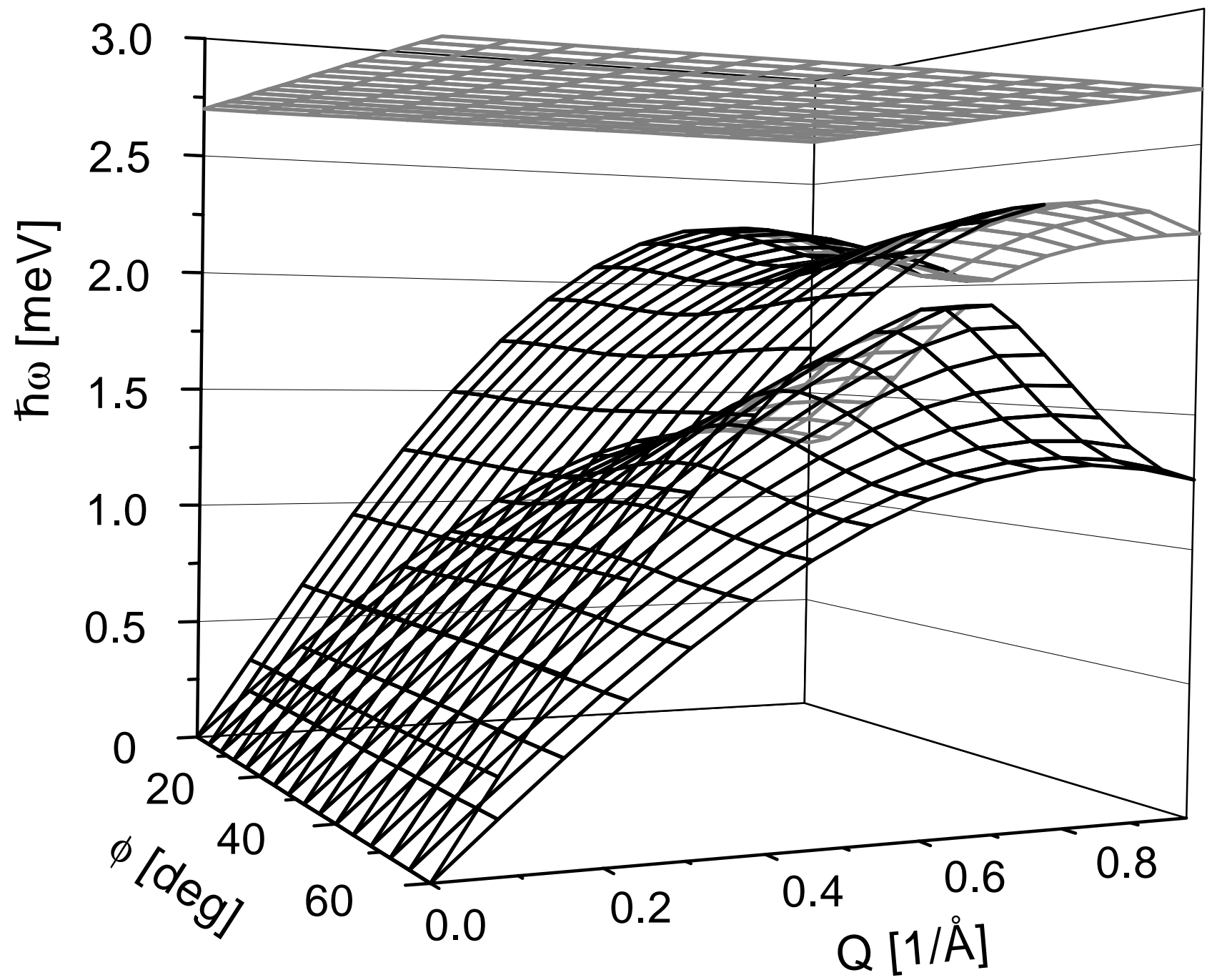
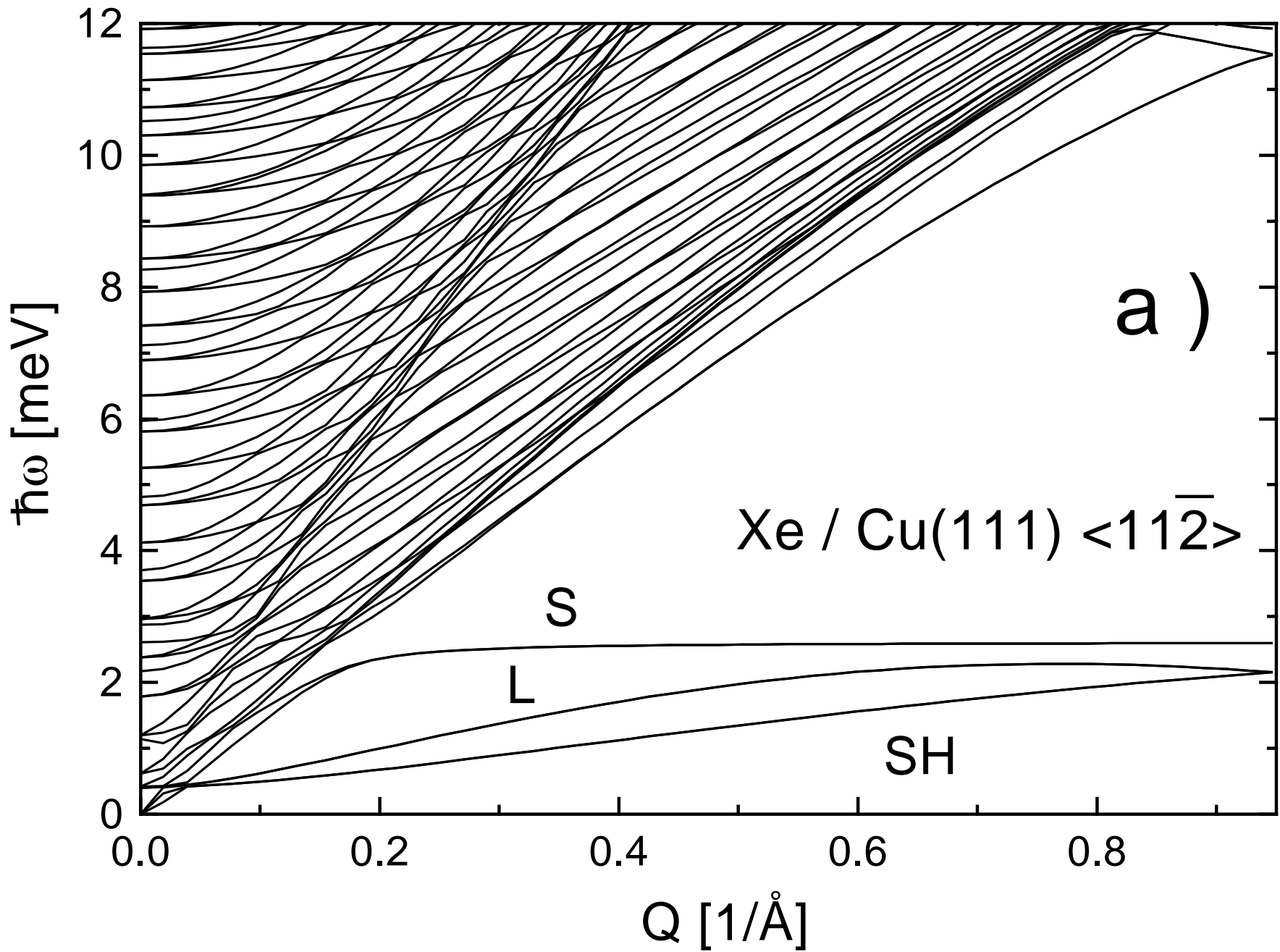
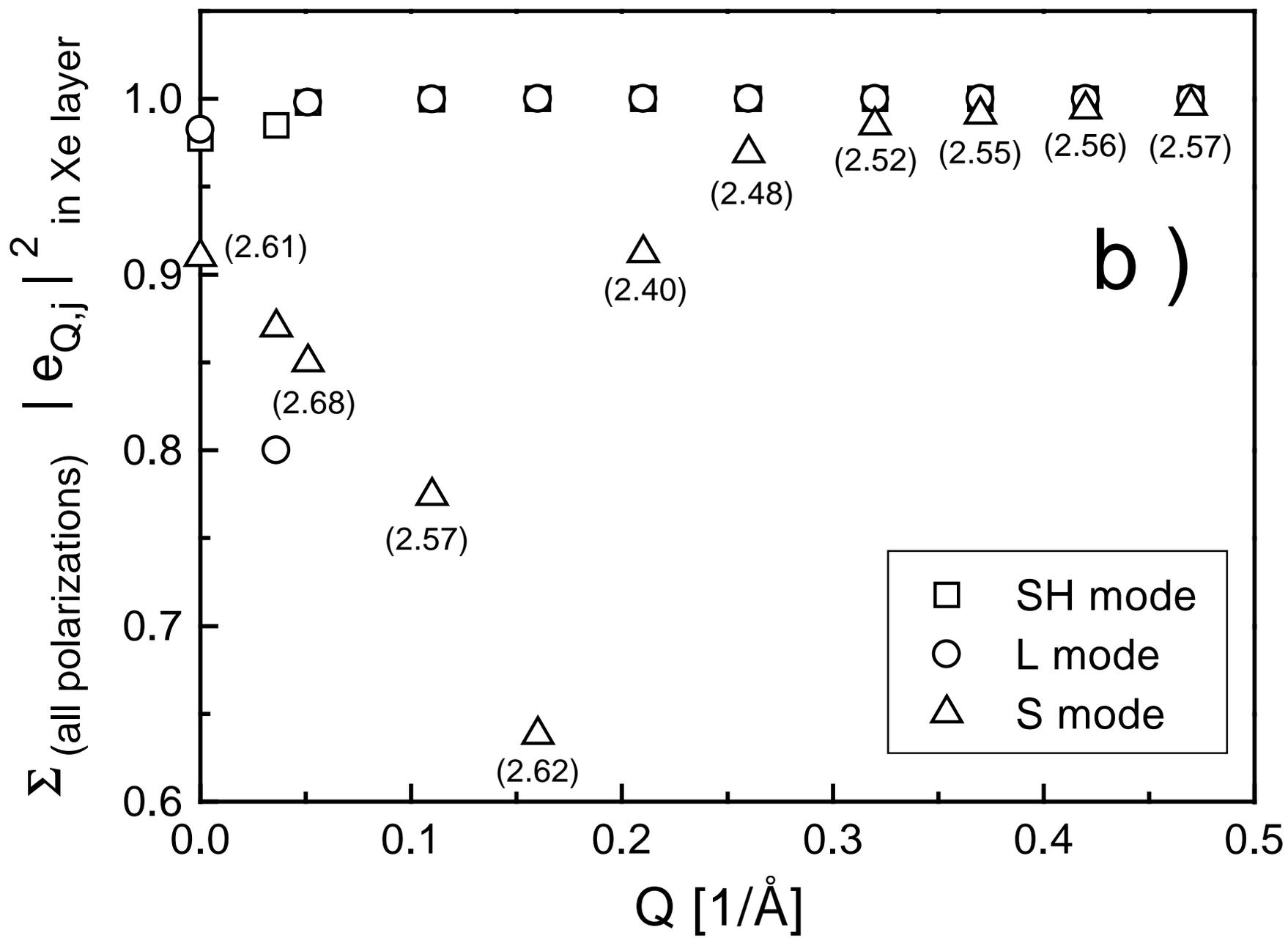
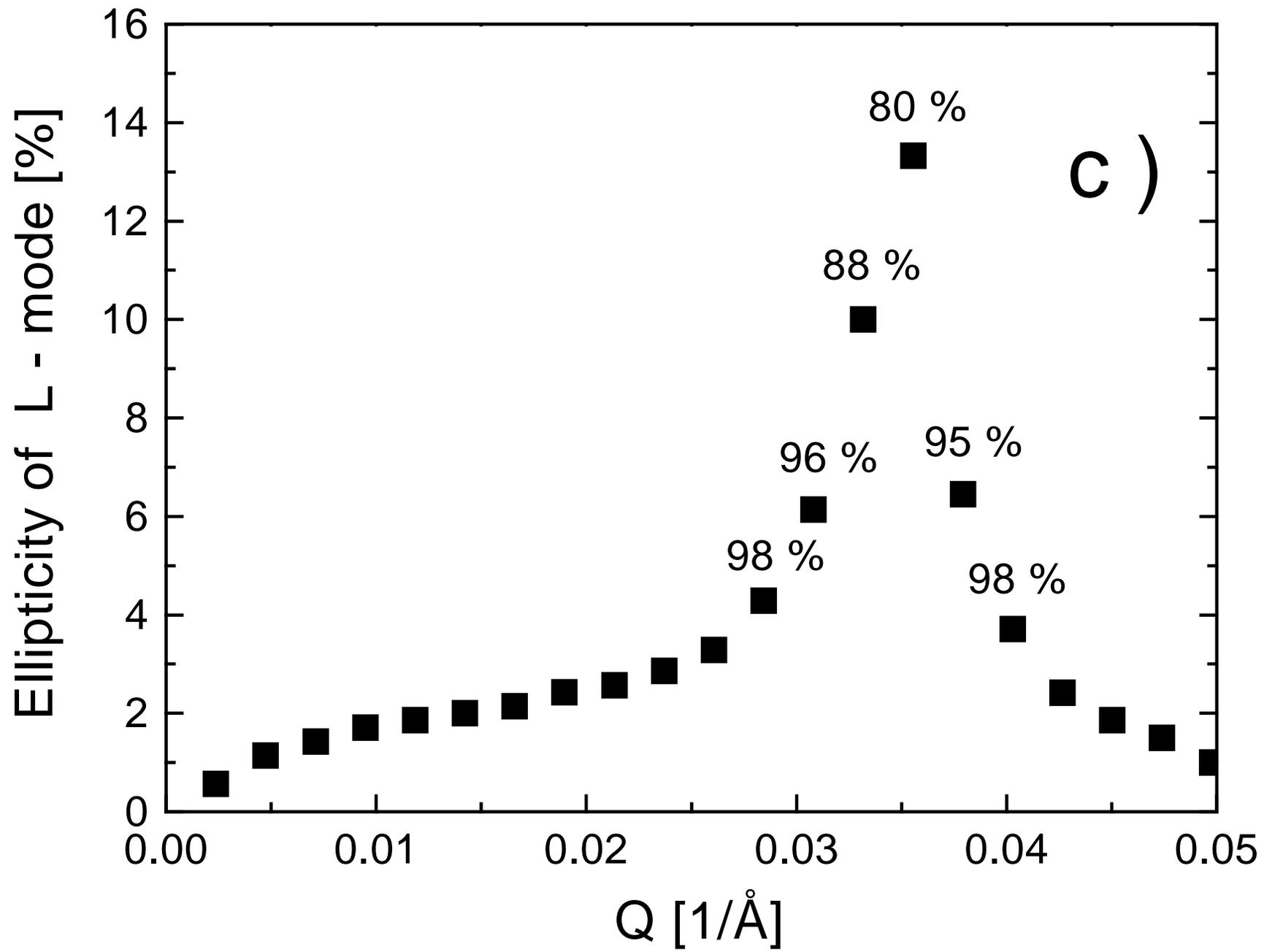
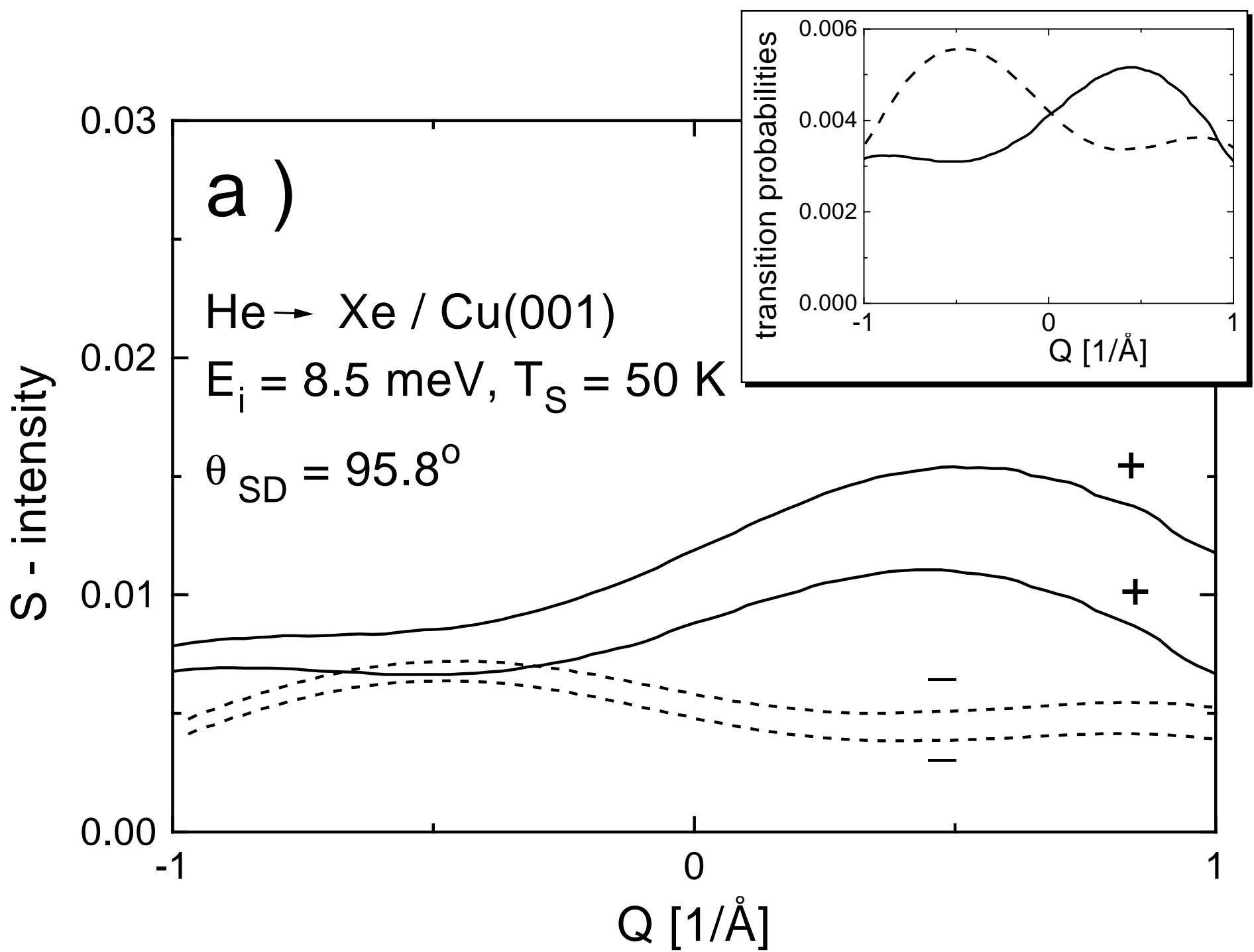
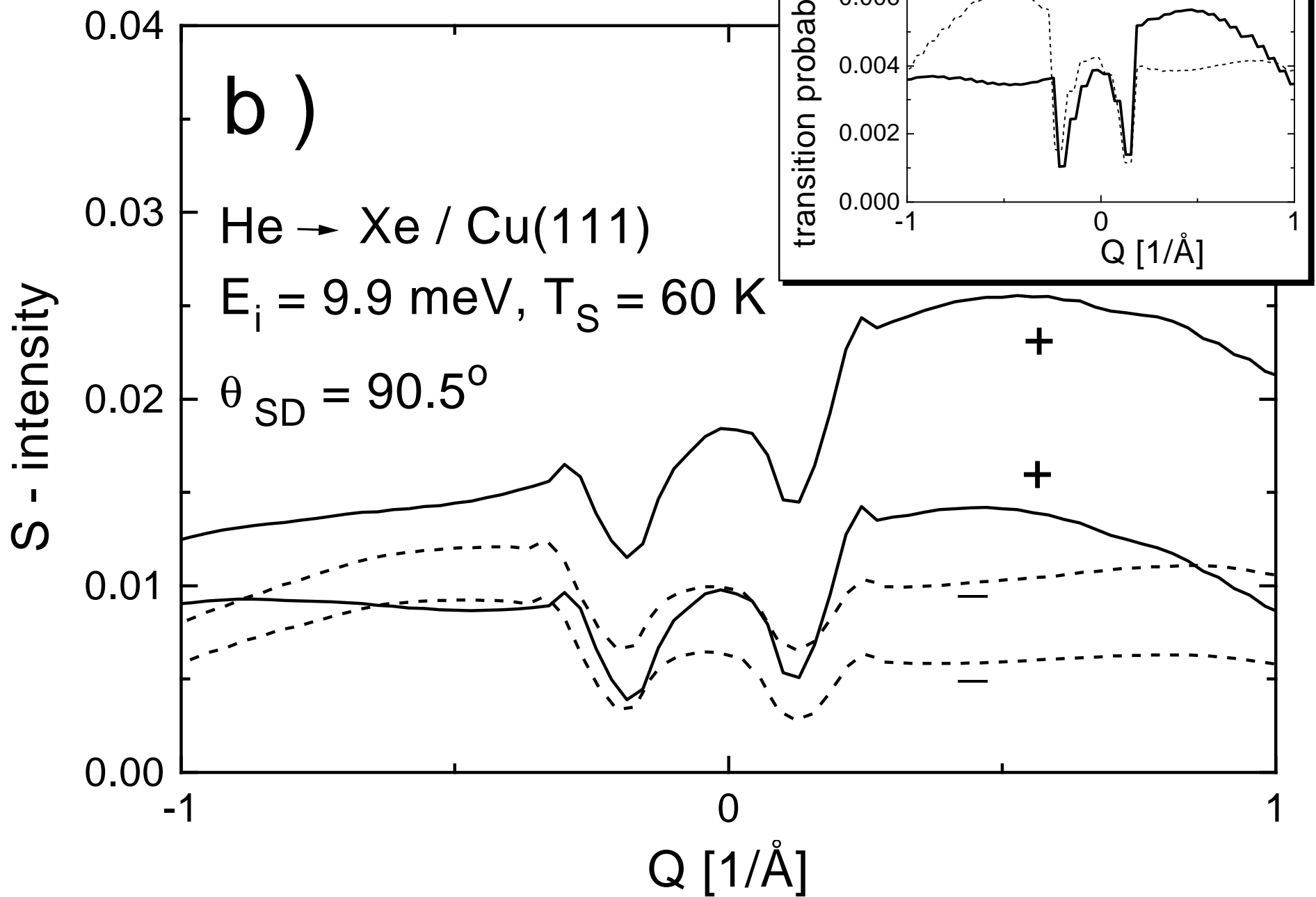
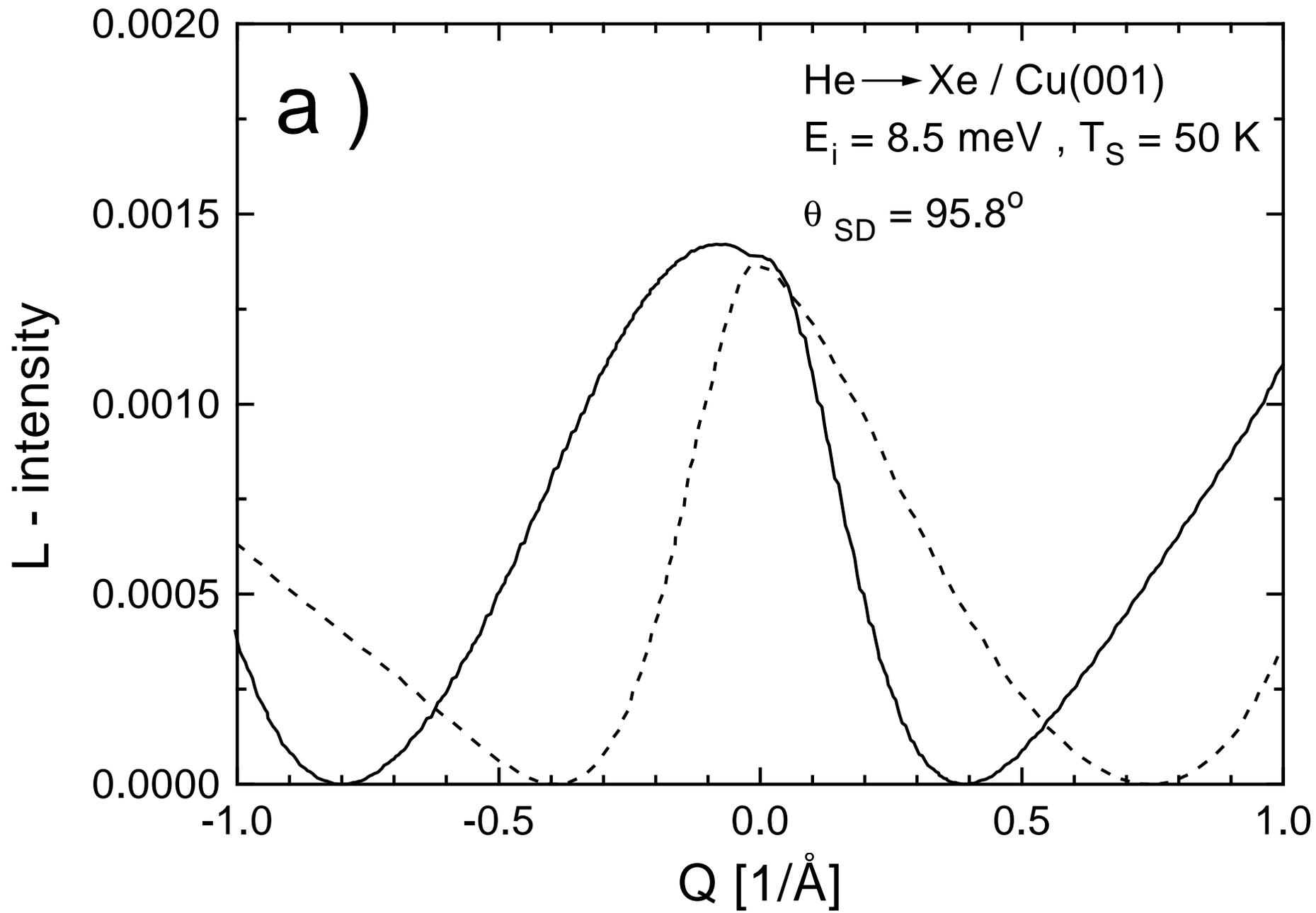
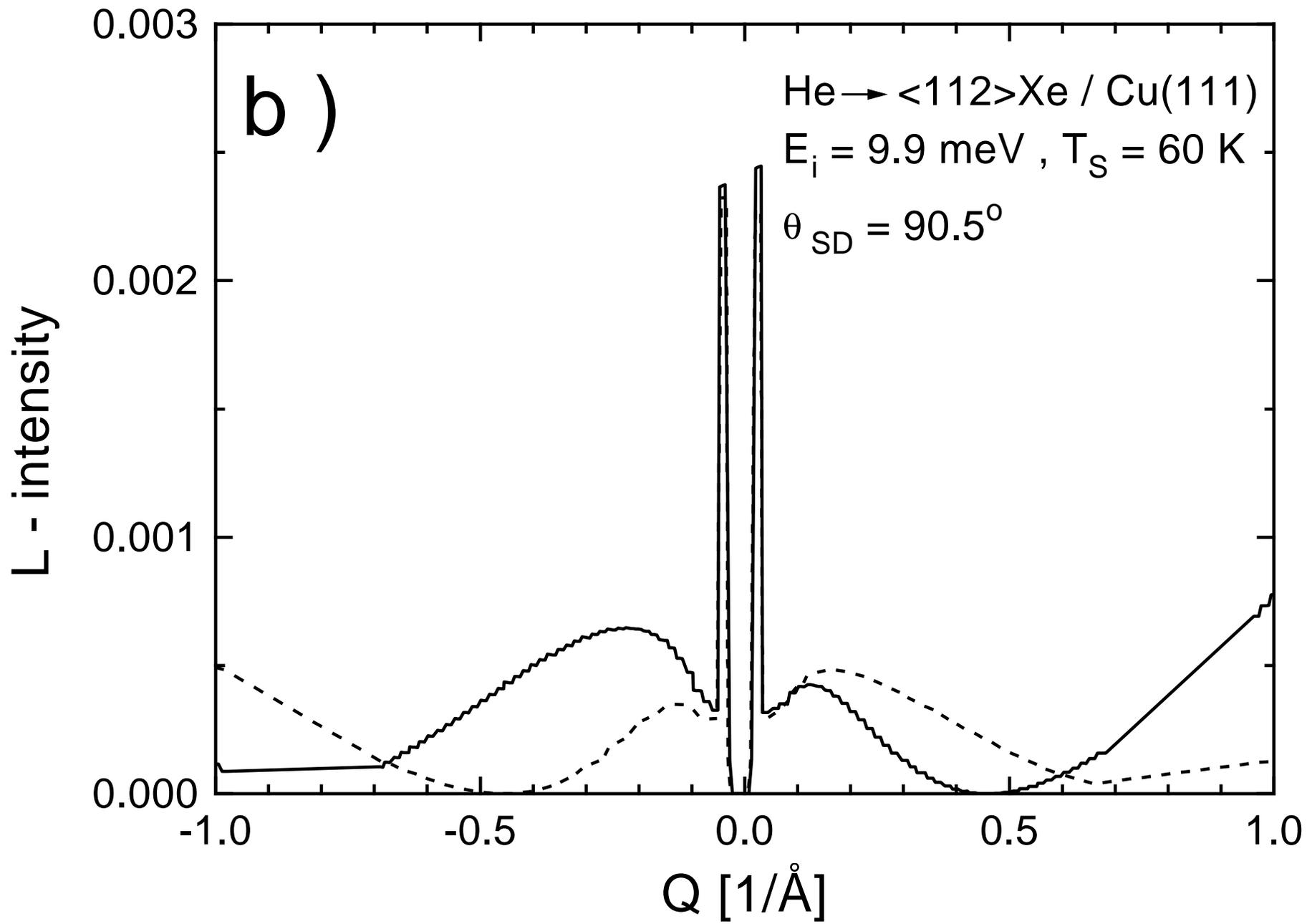
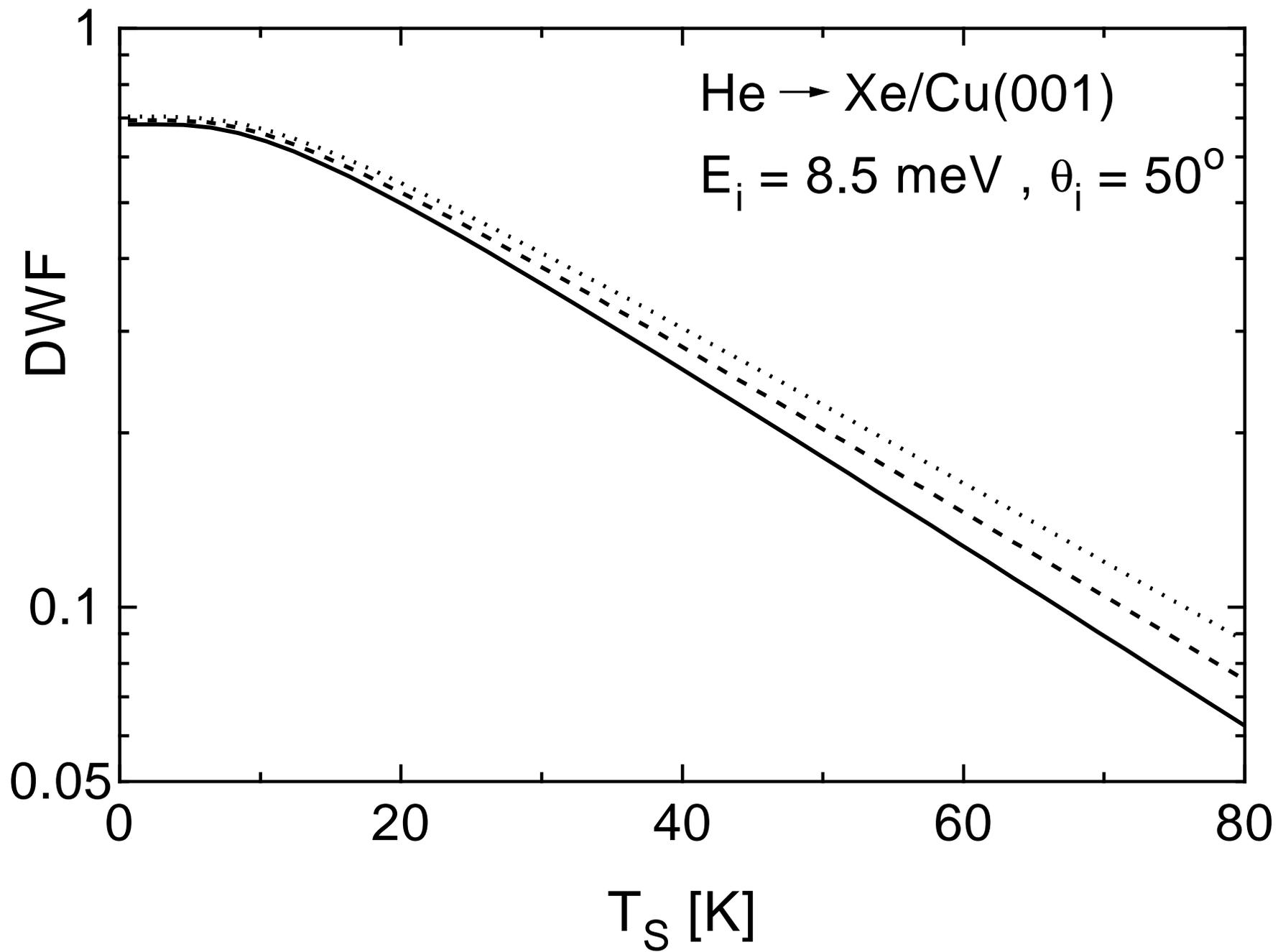
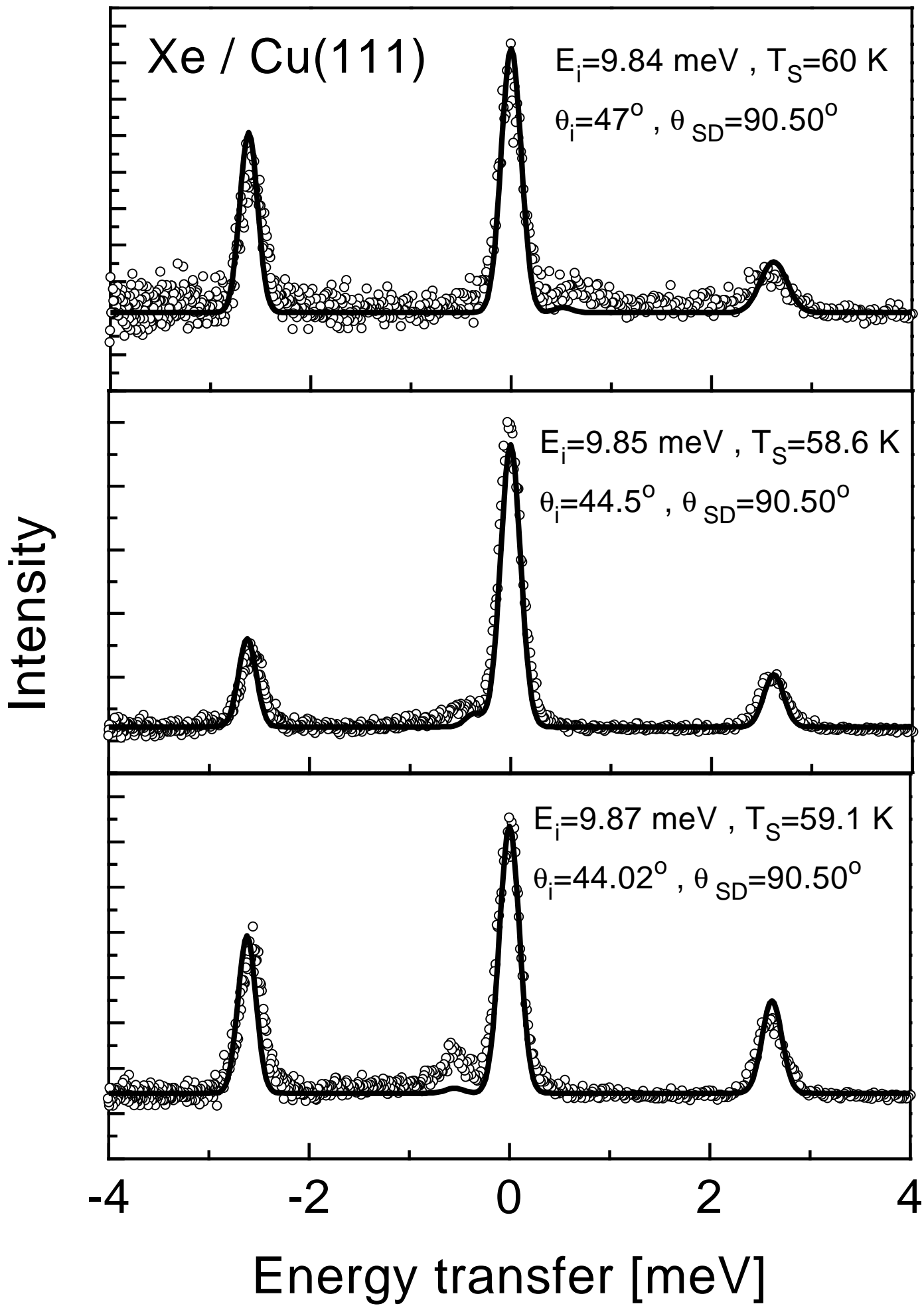
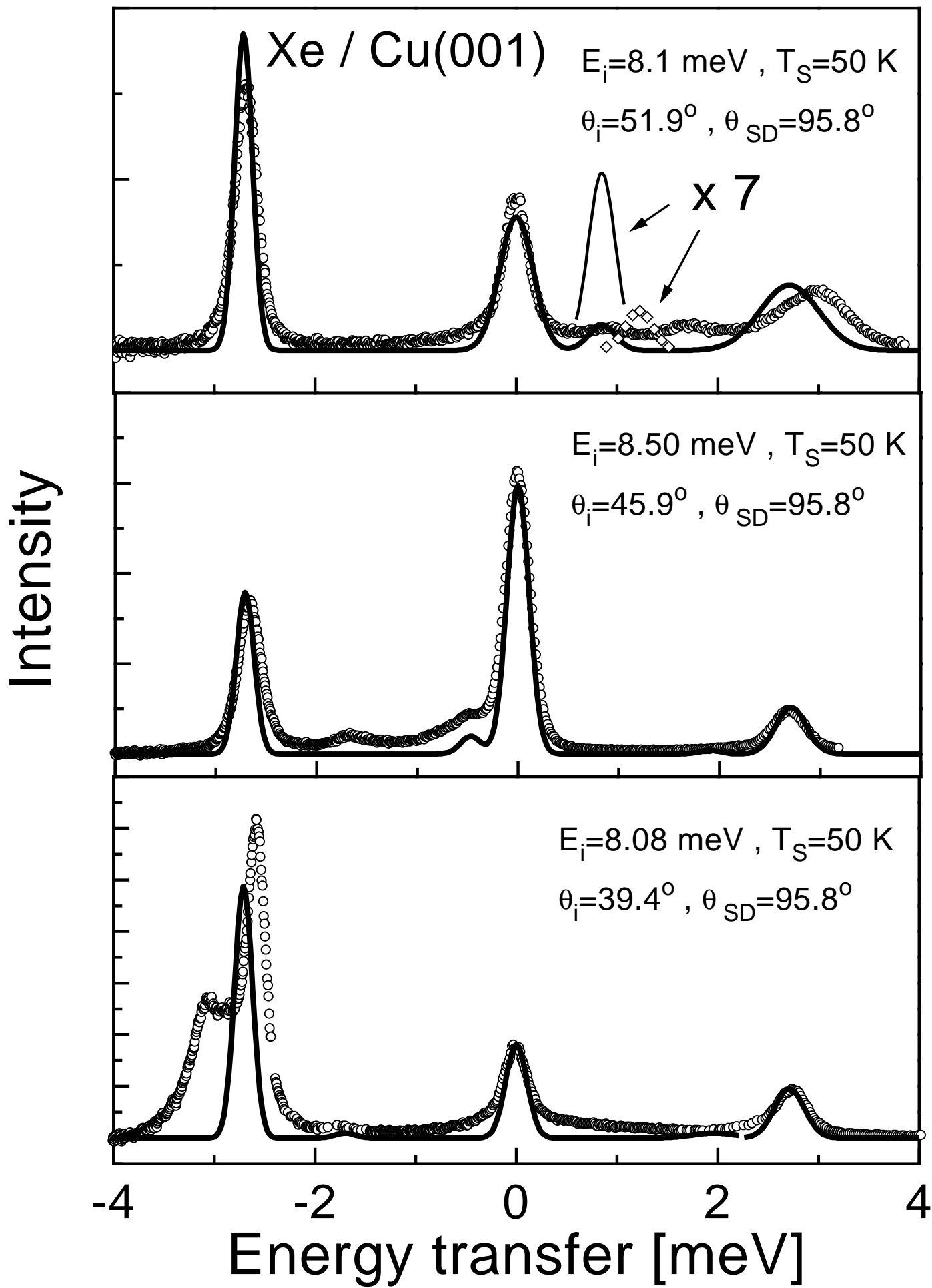
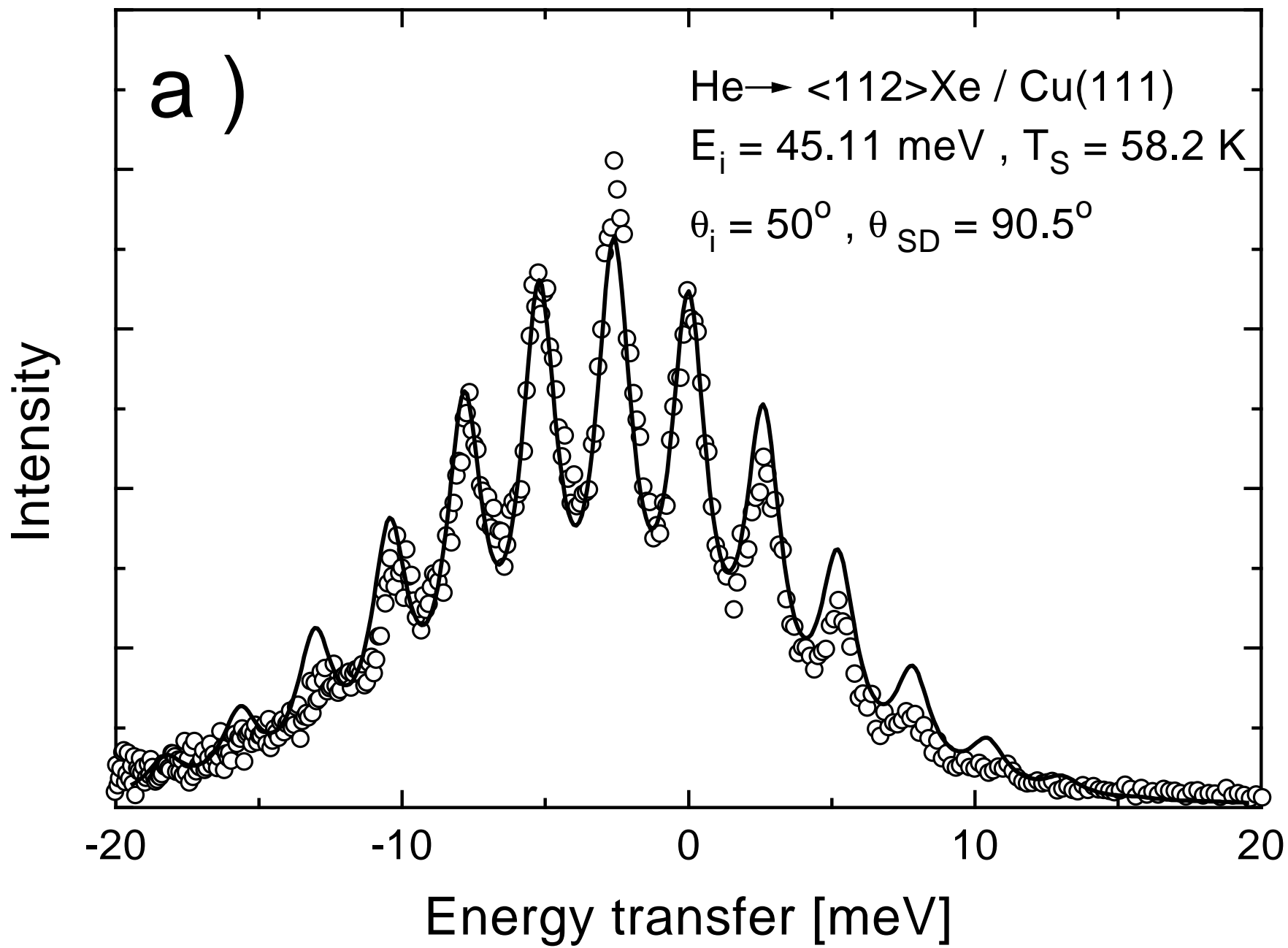
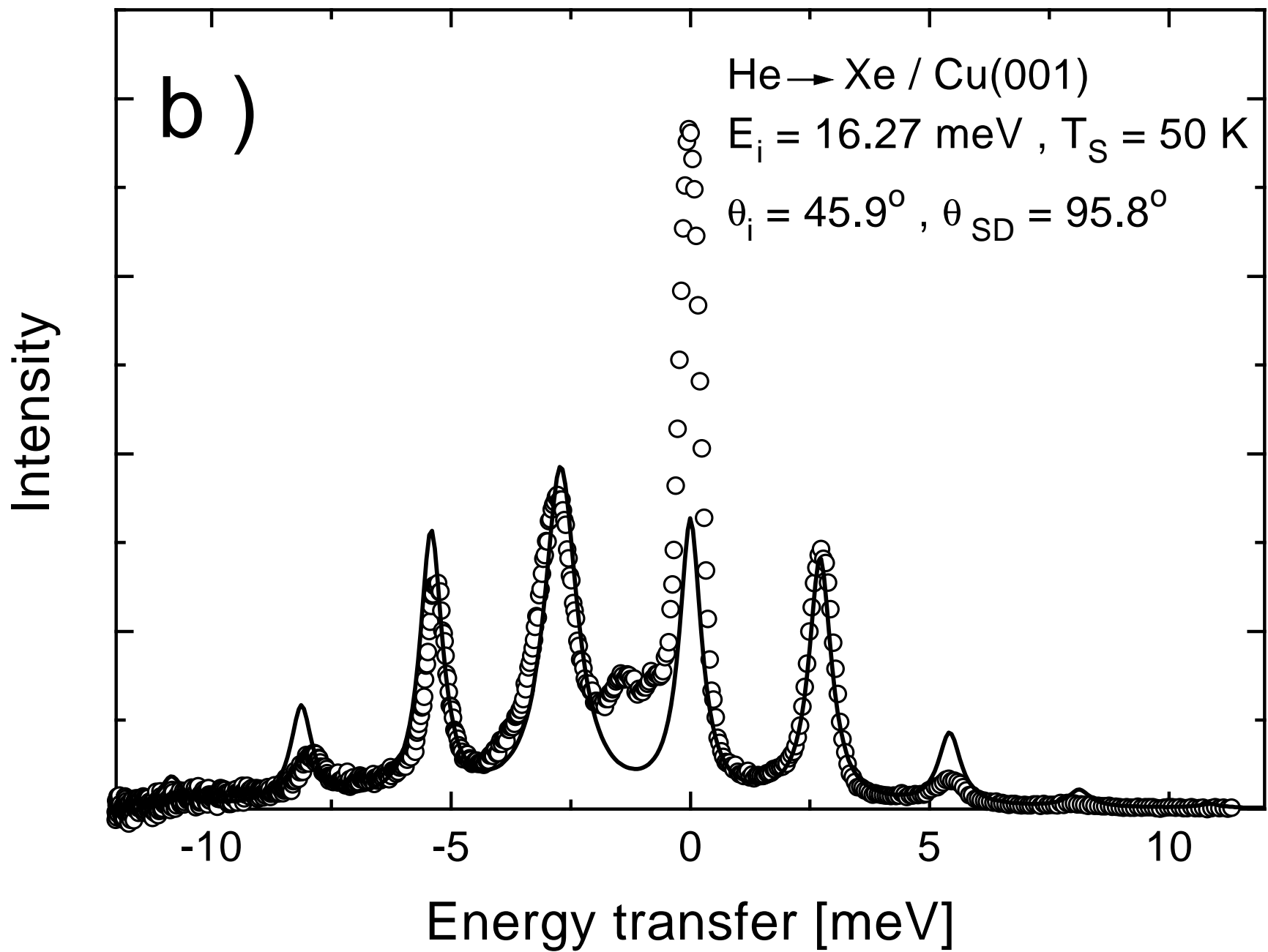
\end{document}